\documentclass[a4paper]{JHEP3i}

\usepackage{graphicx}
\usepackage{amsmath}
\usepackage{cite}
\usepackage{multirow}
\usepackage{slashed}

\def\eqref#1{(\ref{#1})}
\def\text{\rm }

\def\I{\mathcal{I}}

\def\M{\mathcal{M}}
\def\N{\mathcal{N}}
\def\O{\mathcal{O}}

\def\nn{\nonumber \\ }

\newcommand{\GOLEMVC}{{\textsc{go\-lem95C}}}
\newcommand{\QGRAF}{{\sc{QGRAF}}}
\newcommand{\FORM}{{\sc{FORM}}}
\newcommand{\SPINNEY}{{\sc{Spin\-ney}}}
\newcommand{\HAGGIES}{{\sc{Hag\-gies}}}
\newcommand{\SAMURAI}{{\sc{Sa\-mu\-rai}}}

\newcommand{\ninja}{{\sc{Nin\-ja}}}

\newcommand{\bcen}{\begin{center}}
\newcommand{\ecen}{\end{center}}
\def\url#1{\texttt{#1}}

\def\Gosam{{{\sc GoSam}}}
\def\gosam{{{\sc GoSam}}}

\def\blackhat{{{\sc BlackHat}}}

\def\njet{{{\sc NJet}}}

\def\samurai{{{\sc Samurai}}}

\def\cuttools{{{\sc CutTools}}}
\def\C++{{{\sc c++}}}

\def\QCDLoop{{{\sc QCDLoop}}}
\def\OneLoop{{{\sc OneLoop}}}
\def\Golem{{{\sc Golem95C}}}

\def\Ninja{{{\sc Ninja}}}

\def\QCDLoop{{{\sc QCDLoop}}}

\newcommand{\beq}{\begin{equation}}
\newcommand{\eeq}{\end{equation}}
\newcommand{\bqa}{\begin{eqnarray}}
\newcommand{\eqa}{\end{eqnarray}}

\def\spa#1.#2{\langle#1\,#2\rangle}
\def\spb#1.#2{[#1\,#2]}

\def\spab#1.#2.#3{\langle\mskip-1mu{#1}
                  | #2 | {#3}]}

\def\spba#1.#2.#3{[\mskip-1mu{#1}
                  | #2 | {#3}\rangle}

\def\spbb#1.#2.#3.#4{[\mskip-1mu{#1}
                     | {#2} \ {#3} | {#4}]}

\def\spaa#1.#2.#3.#4{\langle\mskip-1mu{#1}
                     | {#2} \ {#3} | {#4}\rangle}

\newcommand{\bite}{\begin{itemize}}
\newcommand{\eite}{\end{itemize}}

\newcommand{\bea}{\begin{eqnarray}}
\newcommand{\eea}{\end{eqnarray}}
\newcommand{\bean}{\begin{eqnarray*}}
\newcommand{\eean}{\end{eqnarray*}}

\title{Multi-leg One-loop Massive Amplitudes from Integrand Reduction via Laurent Expansion}

\author{Hans van Deurzen \\
Max-Planck Insitut f\"ur Physik, F\"ohringer Ring, 6, D-80805 M\"unchen, Germany \\
E-mail: \email{hdeurzen@mpp.mpg.de}
}

\author{Gionata Luisoni \\
Max-Planck Insitut f\"ur Physik, F\"ohringer Ring, 6, D-80805 M\"unchen, Germany \\
E-mail: \email{luisonig@mpp.mpg.de}
}

\author{Pierpaolo Mastrolia \\
Max-Planck Insitut f\"ur Physik, F\"ohringer Ring, 6, D-80805 M\"unchen, Germany \\
Dipartimento di Fisica e Astronomia, Universit\`a di Padova, and INFN Sezione di Padova,
via Marzolo 8, 35131 Padova, Italy\\
E-mail: \email{pierpaolo.mastrolia@cern.ch}
}

\author{Edoardo Mirabella \\
Max-Planck Insitut f\"ur Physik, F\"ohringer Ring, 6, D-80805 M\"unchen, Germany \\
E-mail: \email{mirabell@mpp.mpg.de}
}

\author{Giovanni Ossola \\
Physics Department, New York City College of Technology, The
City University of New York,
300 Jay Street Brooklyn, NY 11201, USA; \\
The Graduate School and University Center, The City University of New York,
365 Fifth Avenue, New York, NY 10016, USA\\
E-mail: \email{GOssola@citytech.cuny.edu}
}

\author{Tiziano Peraro \\
Max-Planck Insitut f\"ur Physik, F\"ohringer Ring, 6, D-80805 M\"unchen, Germany \\
E-mail: \email{peraro@mpp.mpg.de}
}

\abstract{We present the application of a novel reduction technique
  for one-loop scattering amplitudes based on the combination of
  the integrand reduction and Laurent expansion. 
 We describe the general features of its implementation in the
 computer code {\sc Ninja}, and its interface to {\sc GoSam}.
We apply the new reduction to a series of selected processes involving massive particles,
from six to eight legs.}

\begin{document}

\section{Introduction} \label{sec:Intro}

Scattering amplitudes in quantum field theories are analytic
functions of the kinematic variables of the interacting particles, 
hence they can be determined by studying the structure of their singularities. 

The multi-particle factorization properties of the amplitudes become transparent when internal particles go on their mass-shell~\cite{Bern:1994zx,Cachazo:2004kj}.  These
configurations correspond to poles of the amplitude and the investigation of the  general structure of the residues corresponding to  multi-particle factorization channel
turns out to be of particular interest.  Indeed the  knowledge of such structure has been fundamental for 
discovering new relations fulfilled by scattering amplitudes, such as the BCFW  recurrence relation~\cite{Britto:2004ap} and its 
link to the leading singularity of one-loop amplitudes~\cite{Britto:2004nc}.
The systematic classification of the residues, for all the poles corresponding to the quadruple, triple, double, and single cuts, has been achieved, in four dimensions, by employing integrand-reduction methods~\cite{delAguila:2004nf, Ossola:2006us}. The latter  led to the OPP integrand-decomposition formula for  one-loop integrals~\cite{Ossola:2006us}, which 
allows one to write each residue as a linear combination of  process-independent polynomials multiplied by process-dependent coefficients.

These results provided a deeper understanding of the structure of scattering amplitudes and have shown the underlying simplicity beneath the rich mathematical structure of quantum field theory. 
Moreover, they provided the theoretical framework for the development of efficient computational algorithms for one-loop calculations in perturbation theory, which have been implemented in various 
automated codes~\cite{Berger:2008sj,Giele:2008bc,Badger:2010nx,Bevilacqua:2011xh, Hirschi:2011pa, Cullen:2011ac,Agrawal:2011tm,Cascioli:2011va, Actis:2012qn,Badger:2012pg} 
improving the state-of-the art of the predictions at the next-to-leading order accuracy~\cite{Bevilacqua:2009zn, Lazopoulos:2008ex, Giele:2009ui, vanHameren:2009vq,Berger:2009zg,
Berger:2009ep,Ellis:2009zw,KeithEllis:2009bu, Berger:2010vm, Bevilacqua:2010qb,Denner:2010jp,Melia:2011dw,Bevilacqua:2011aa,Greiner:2011mp,Campanario:2011ud,Denner:2012yc, Greiner:2012im,
Cullen:2012eh,Campanario:2012bh,Bevilacqua:2012em,Gehrmann:2013aga,Campanario:2013qba,Campanario:2013mga,Bevilacqua:2013taa, Greiner:2013gca,vanDeurzen:2013rv,Gehrmann:2013bga,Cullen:2013saa,vanDeurzen:2013xla, Campanario:2013fsa, Campanario:2013gea, Dolan:2013rja, Badger:2013ava, Heinrich:2013qaa}.

Recently, the classification of the structure of the residues has been obtained in a more general and elegant form within the framework of 
multivariate polynomial division and algebraic geometry~\cite{Mastrolia:2012an,Zhang:2012ce}.
The use of these techniques proved that the integrand decomposition, originally formulated for one-loop amplitudes, is applicable
at any order in perturbation theory, irrespective of the complexity of the topology of the diagrams involved.
An iterative integrand-recursion formula, based on successive divisions of the numerators modulo the 
Gr\"obner basis of the ideals generated by the cut denominators, can provide the form of the residues at all multi-particle poles for arbitrary amplitudes, independently of the number of loops.
Extensions of the integrand reduction method beyond one-loop, initiated in~\cite{Mastrolia:2011pr,Badger:2012dp} and then systematized within the language of algebraic 
geometry~\cite{Mastrolia:2012an,Zhang:2012ce} have recently become the topic of several studies~\cite{Kleiss:2012yv, Badger:2012dv,Feng:2012bm, Mastrolia:2012wf, Huang:2013kh, 
Mastrolia:2013kca}, thus providing a new direction in the study of multi-loop amplitudes.

In the context of the integrand reduction, the process-dependent  coefficients can be numerically determined by 
solving a system of algebraic equations that are obtained by  evaluating  the numerator of the integrand at explicit values of the loop momentum~\cite{Ossola:2006us}.  The system
becomes triangular if one evaluates the numerator at the multiple cuts, i.e. at the set of complex values of the integration momentum for which a given set of propagators 
vanish. The extraction of all coefficients via  polynomial fitting has been implemented  in publicly available codes performing integrand decomposition, such as \cuttools~\cite{Ossola:2007ax} and \samurai~\cite{Mastrolia:2010nb}. These algorithms do not require any  specific recipe for the generation of the numerator function, which can be performed by using traditional Feynman diagrams, by means of recursive relations, or by gluing tree-level sub-amplitudes, as in unitarity-based methods.

The code {\cuttools} implements  the four-dimensional  integrand-reduction algorithm~\cite{Ossola:2007bb, Mastrolia:2008jb,Ossola:2008xq}, in which the cut-constructible term and the rational term are necessarily computed separately.  The latter escapes  four-dimensional integrand reduction 
and has to be computed by means of  other methods, e.g.  \emph{ad hoc} tree-level Feynman rules~\cite{Ossola:2008xq,Draggiotis:2009yb,Garzelli:2009is, Garzelli:2010qm,  Garzelli:2010fq, Shao:2011tg, Shao:2012ja,Page:2013xla}.

Significant improvements were achieved with the $d$-dimensional extension of integrand-reduction methods~\cite{Ellis:2007br, Giele:2008ve,Ellis:2008ir}, which expose a richer polynomial structure of the integrand and allows for the combined determination of both cut-constructible and rational terms at once. 
This idea of performing unitarity-cuts in $d$-dimension was the basis for the development of {\samurai}, which extends  the OPP polynomial structures to include an explicit dependence on the $(d-4)$-dimensional parameter needed for the automated computation of the full rational term. Moreover, it includes  the parametrization of the residue of the quintuple-cut~\cite{Melnikov:2010iu} and implements the numerical sampling via Discrete Fourier Transform~\cite{Mastrolia:2008jb}.

The integrand decomposition was originally developed for renormalizable gauge theories,
where, at one-loop, the rank of the numerator cannot be larger than the number of external legs.
The reduction of diagrams where the rank can be higher, as 
required for example when computing  $pp \to H +2,3$ jets in gluon fusion in the large top-mass limit~\cite{vanDeurzen:2013rv, Cullen:2013saa}, 
demands an extension of the algorithm  to  accommodate the richer monomial structures of  the residues. This extension has been implemented in  {\samurai}, together with the 
corresponding sampling required to fit all the coefficients~\cite{Mastrolia:2012du, Mastrolia:2012bu, vanDeurzen:2013pja}.

More recently, elaborating on the the techniques proposed in~\cite{Forde:2007mi,Badger:2008cm}, a different approach to the $d$-dimensional integrand-reduction method was proposed~\cite{Mastrolia:2012bu}. 
The key point of this method is to extract the coefficients more efficiently by performing a Laurent expansion of the integrand. The method is general and relies only on the knowledge of the explicit dependence  of the numerator  on the loop momentum.

In general, when the multiple-cut conditions do not fully constrain the loop momentum, the
solutions are still functions of some free parameters, possibly the components of the momentum which are not frozen by the cut conditions.
The integrand-reduction algorithms  implemented in the codes~\cite{Ossola:2007ax,Mastrolia:2010nb} require to solve
a  system of equations obtained by sampling on those free parameters. The system is triangular thus  the coefficients of a given residue can be computed  only 
after subtracting all the non-vanishing contributions coming from higher-point residues.

The key observation suggested in Ref.\cite{Mastrolia:2012bu} is that 
the reduction algorithm can be simplified by exploiting the universal structure of the residues, hence of their asymptotic expansion. Indeed, by performing a
\emph{Laurent expansion} with respect to one of the free parameters
which appear in the solutions of the cut,
both the integrand and the subtraction terms exhibit the same polynomial behavior of the residue.  
Moreover, the contributions coming
from the subtracted terms can be implemented as \emph{corrections at
  the coefficient level}, hence replacing the subtractions at the
integrand level of the original algorithm. 
The parametric form of these corrections can be computed once and for all, in terms of a
subset of the higher-point coefficients. 
This method significantly reduces the number
of coefficients entering each subtracted term, in particular boxes and pentagons 
decouple from the computation of lower-points coefficients.

If either the analytic expression of the integrand or the tensor structure of
the numerator is known, this procedure can be implemented in a
semi-numerical algorithm. Indeed, the coefficients of the Laurent
expansion of a rational function can be computed, either analytically
or numerically, by performing a {\it polynomial division} between the
numerator and the (uncut) denominators. 
\bigskip

The scope of the present paper is to review the main features of the novel reduction algorithm and demonstrate its performance on a selection of challenging calculations of scattering amplitudes with massive bosons and quarks, involving {\it six}, {\it seven}, and {\it eight} particles.
The integrand-reduction via Laurent expansion has been implemented in the {\sc c++} library {\sc Ninja}~\cite{Peraro:2013oja, Peraro:2014cba}, and interfaced to the {\Gosam} framework~\cite{Cullen:2011ac} for the evaluation of virtual one-loop scattering amplitudes.
The cleaner and lighter system-solving strategy, which deals with a diagonal system instead of a triangular one, and which substitutes 
the polynomial subtractions with coefficients corrections, turns into net gains in terms of both numerical accuracy and computing speed.
We recall that the new library has been recently used  in the evaluation of NLO QCD corrections to $p p \to t {\bar t} H j$~\cite{vanDeurzen:2013xla}.

The paper is organized as follows.
In Section~\ref{sec:Laurent}, we discuss the theoretical foundations
of the integrand decomposition via Laurent
expansion, and  its implementation in an
algorithm for the reduction of one-loop amplitudes. 
The description of the interface between {\Ninja} and {\gosam} for
automated one-loop calculation is discussed in
Section~\ref{sec:Ninja}.
Section~\ref{sec:Precision} is devoted to a detailed study
of the precision and of the computational performance of the novel
framework, which shows a significant improvement with respect to the
standard algorithms.
Applications of the  {\Gosam}+{\Ninja} framework
to the evaluation  of NLO QCD virtual correction to several multi-leg
massive processes are shown in
Section~\ref{sec:Applications}.

\section{Reduction Algorithm -- Integrand Reduction via Laurent Expansion} \label{sec:Laurent}

In this section we describe the \emph{Laurent-expansion method} for
the integrand reduction of one-loop amplitudes as implemented in the
{\sc C++} library {\sc Ninja}.

\subsection{Integrand and Integral decomposition}
\label{sec:decomposition}

An $n$-point one-loop amplitude can be written as a linear combination
of contributions $\M_{1\cdots n}$ of the form
\begin{equation}
  \M_{1\cdots n} = \int d^d \bar q \; \I_{1\cdots n} = \int d^d \bar q\, \frac{\N(\bar q)}{D_1\cdots D_n}
\end{equation}
where $\N(\bar q)$ is a process-dependent polynomial numerator in the
components of the $d=(4-2 \epsilon)$-dimensional  
loop momentum $\bar q$. The latter can be decomposed as follows,
\bea
\bar {\slashed q} = {\slashed q} + {\slashed \mu} \ ,  \qquad  \bar q^2 = q^2 -\mu^2 \ ,
\eea
in terms of its $4$-dimensional component, $q$, and $\mu^2$ which encodes 
its  $(-2\epsilon)$-dimensional components. The denominators $D_i$
are quadratic polynomials in $\bar q$ and correspond to Feynman loop
propagators,
\begin{equation}
  \label{eq:loopdenom}
  D_i = (\bar q + p_i)^2-m_i^2.
\end{equation}
Every one-loop integrand in $d$ dimensions can be
decomposed as \cite{Ossola:2006us,Ellis:2007br}
\begin{equation}
  \label{eq:integranddec}
  \I_{1\ldots n} \equiv \frac{\N_{1\cdots n}}{D_1\ldots D_n} = \sum_{k=1}^5\sum_{\{i_1,\cdots, i_k\}}\frac{\Delta_{i_1\cdots i _k}}{D_{i_1}\cdots D_{i_k}},
\end{equation}
where the $\Delta_{i_1\cdots i_k}$ are irreducible polynomial
residues, i.e.\ polynomials which do not contain any term proportional
to the corresponding loop denominators $D_{i_1},\ldots,D_{i_k}$.  The
second sum in Eq.~\eqref{eq:integranddec} runs over  all
unordered
selections without repetition of the $k$ indices $\{i_1,\cdots,
i_k\}$.

For any set of denominators $D_{i_1},\ldots,D_{i_k}$ with $k\leq 5$,
one can choose a $4$-dimensional basis of momenta $\mathcal{E}=\{e_1,e_2,e_3,e_4\}$
which satisfies the following  normalization conditions
 \begin{align}
e_1\cdot e_2 =1, \qquad e_3^2 = e_4^2 = \delta_{k4},  \qquad     e_3 \cdot e_4 =-(1-\delta_{k4}),
\end{align}
while all the other scalar products vanish.  In addition,
for $k=4$ we choose the basis such that $e_4$ is orthogonal to the
external legs of the sub-diagram identified by the set of denominators
in consideration.  Similarly, for $k=2,3$ we choose both $e_3$
and $e_4$ to be orthogonal to the external legs of the corresponding
sub-diagram.  With this choices of momentum basis, the numerator and
the denominators can be written 
as polynomials in the coordinates $\mathbf{z} \equiv (z_1,z_2,z_3,z_4,z_5)
= (x_1,x_2,x_3,x_4,\mu^2)$. The variables $x_i$ are the
components of $q$ with respect to the basis $\mathcal{E}$, 
\begin{equation}
  q^\nu  = -p^\nu_{i_1} +  x_1 \ e^\nu_1 + x_2 \ e^\nu_2 + x_3 \ e^\nu_3 + x_4 \ e^\nu_4 \ .%\qquad \bar q^2 = q^2-\mu^2.
\end{equation}
More explicitly, the numerator is a polynomial in the components of
$q$ and $\mu^2$
\begin{equation}
  \N(\bar q) = \N(q,\mu^2) = \N(x_1,x_2,x_3,x_4,\mu^2) = \N(\mathbf{z}).
\end{equation}
The coordinates $x_i$ can also be written as scalar
products,
% \begin{alignat}{2}
%   & x_1 = ((q+p_{i_1})\cdot e_2), \qquad & & x_3 = -((q+p_{i_1})\cdot e_4)(1-\delta_{k4}) + ((q+p_{i_1})\cdot e_3)\delta_{k4} \nn
%   & x_2 = ((q+p_{i_1})\cdot e_1), \qquad & & x_4 = -((q+p_{i_1})\cdot e_3)(1-\delta_{k4}) + ((q+p_{i_1})\cdot e_4)\delta_{k4} 
% \end{alignat}
\begin{align}
  x_1 & = (q+p_{i_1})\cdot e_2 \nn
  x_2 & = (q+p_{i_1})\cdot e_1 \nn
  x_3 & = -((q+p_{i_1})\cdot e_4)(1-\delta_{k4}) + ((q+p_{i_1})\cdot e_3)\delta_{k4}\nn
  x_4 & = -((q+p_{i_1})\cdot e_3)(1-\delta_{k4}) + ((q+p_{i_1})\cdot e_4)\delta_{k4}.
\end{align}

With these definitions, one can show \cite{Ossola:2006us,Ellis:2007br,Mastrolia:2012an} that the most general
parametric form of a residue in a renormalizable theory is
\begin{align}
  \Delta_{i_1 i_2 i_3 i_4 i_5} ={}& c^{(i_1 i_2 i_3 i_4 i_5)}_0\, \mu^2 \nonumber \\[1.0ex]
  \Delta_{i_1 i_2 i_3 i_4} ={}& c^{(i_1 i_2 i_3 i_4)}_0 + c^{(i_1 i_2 i_3 i_4)}_1 x_4 + \mu^2 \left ( c^{(i_1 i_2 i_3 i_4)}_2 + c^{(i_1 i_2 i_3 i_4)}_3 x_4 + \mu^2 c^{(i_1 i_2 i_3 i_4)}_4 \right )\nn
  {\Delta}_{i_1 i_2 i_3} ={}& c^{(i_1 i_2 i_3)}_0 + c^{(i_1 i_2 i_3)}_1 x_3 + c^{(i_1 i_2 i_3)}_2 x_3^2 + c^{(i_1 i_2 i_3)}_3 x_3^3
  + c^{(i_1 i_2 i_3)}_4 x_4 + c^{(i_1 i_2 i_3)}_5 x_4^2  \nn
  {}&+ c^{(i_1 i_2 i_3)}_6 x_4^3
+ \mu^2 \left (c^{(i_1 i_2 i_3)}_7 + c^{(i_1 i_2 i_3)}_8 x_3 + c^{(i_1 i_2 i_3)}_9 x_4 \right ) \nonumber \\[1.0ex]
{\Delta}_{i_1 i_2} ={}& c^{(i_1 i_2)}_0  + c^{(i_1 i_2)}_1 x_2 + c^{(i_1 i_2)}_2 x_3 + c^{(i_1 i_2)}_3 x_4
+   c^{(i_1 i_2)}_4 x^2_2 + c^{(i_1 i_2)}_5 x^2_3  \nn
{}&+ c^{(i_1 i_2)}_6 x^2_4 + c^{(i_1 i_2)}_7 x_2 x_3
+ c^{(i_1 i_2)}_8 x_2x_4  + c^{(i_1 i_2)}_9 \mu^2  \nonumber \\[1.0ex]
{\Delta}_{i_1} ={}& c^{(i_1)}_0  + c^{(i_1)}_1 x_1 + c^{(i_1)}_2 x_2 + c^{(i_1)}_3 x_3
+   c^{(i_1)}_4 x_4.
\label{eq:parametricresidues}
\end{align}
This parametric form can also be extended to non-renormalizable and
effective theories, where the rank of the numerator can be larger
than the number of loop propagators \cite{Mastrolia:2012bu}.  
Most of the term appearing in
Eq.~\eqref{eq:parametricresidues} vanish after integration, i.e. 
they are \emph{spurious}. The non-spurious coefficients, instead, 
appear in the final result which expresses the amplitude $\M_{1\cdots n}$ 
as a linear combination of known Master Integrals, 
\begin{align}
  \M_{1\cdots n} = {}&
\sum_{\{i_1, i_2, i_3, i_4\}}\bigg\{
          c_{0}^{ (i_1 i_2 i_3 i_4)} I_{i_1 i_2 i_3 i_4} +  
          c_{4}^{ (i_1 i_2 i_3 i_4)} I_{i_1 i_2 i_3 i_4}[\mu^4] 
\bigg\} \pagebreak[1] \nn
     & +
\sum_{\{i_1, i_2, i_3\}}^{n-1}\bigg\{
          c_{0}^{ (i_1 i_2 i_3)} I_{i_1 i_2 i_3} +
          c_{7}^{ (i_1 i_2 i_3)} I_{i_1 i_2 i_3}[\mu^2]
\bigg\} \pagebreak[1] \nn
     & +
\sum_{\{i_1, i_2\}}\bigg\{
          c_{0}^{ (i_1 i_2)} I_{i_1 i_2} 
        + c_{1}^{ (i_1 i_2)} I_{i_1 i_2}[(q+p_{i_1})\cdot e_2 ] \nn
        & \qquad
        + c_{2}^{ (i_1 i_2)} I_{i_1 i_2}[((q+p_{i_1})\cdot e_2)^2 ]         +  c_{9}^{ (i_1 i_2)} I_{i_1 i_2}[\mu^2]  \bigg\} \pagebreak[1] \nn 
& + 
\sum_{i_1}
      c_{0}^{ (i_1)} I_{i_1},   \label{eq:integraldecomposition}
\end{align}
where
\begin{equation*}
  I_{i_1 \cdots i_k}[\alpha] \equiv \int d^d \bar q {\alpha \over D_{i_1} \cdots D_{i_k} }, 
\qquad
I_{i_1 \cdots i_k} \equiv   I_{i_1 \cdots i_k}[1].
\end{equation*}

The problem of the computation of any one-loop amplitude can therefore
be reduced to the problem of the determination of the coefficients of
the Master Integrals appearing in
Eq.~\eqref{eq:integraldecomposition}, which in turn can be identified
with a subset of the coefficients of the parametric residues in
Eq.~\eqref{eq:parametricresidues}.

\subsection{Scattering amplitudes via Laurent expansion}
In Ref.~\cite{Mastrolia:2012bu}, elaborating on the the techniques
proposed in~\cite{Forde:2007mi,Badger:2008cm}, an improved version of
the integrand-reduction method for one-loop amplitudes was presented.
This method allows, whenever the analytic dependence of the integrand
on the loop momentum is known, to extract the unknown coefficients of
the residues $\Delta_{i_1\cdots i_k}$ by performing a Laurent
expansion of the integrand with respect to one of the free loop
components which are not constrained by the corresponding on-shell
conditions $D_{i_1}=\cdots = D_{i_k}=0$.

Within the original integrand reduction algorithm~\cite{Ossola:2007ax,
  Mastrolia:2008jb, Mastrolia:2010nb}, the determination of these
unknown coefficients requires: {\it i)} to sample the numerator on a finite
subset of the on-shell solutions; {\it ii)} to subtract from the integrand the
non-vanishing contributions coming from higher-point residues; and
{\it iii)} to solve the resulting linear system of equations.

With the Laurent-expansion approach, since in the asymptotic limit
both the integrand and the higher-point subtractions exhibit the same
polynomial behavior as the residue, one can instead identify the
unknown coefficients with the ones of the expansion of the integrand,
corrected by the contributions coming from higher-point residues.  In
other words, the system of equations for the coefficients becomes
diagonal and the subtractions of higher-point contributions can be
implemented as \emph{corrections at the coefficient level} which
replace the subtractions at the integrand level of the original
algorithm.  Because of the universal structure of the residues, the parametric form of these corrections can be computed
once and for all, in terms of a subset of the higher-point
coefficients.  This also allows to significantly reduce the number of
coefficients entering in each subtraction.  For instance, box and
pentagons do not affect at all the computation of lower-points
residues.  In the followings, we describe in more detail how to
determine the needed coefficients of each residue.

\paragraph{Pentagons}
Pentagons contributions are spurious, i.e.\ they do not appear in the
integrated result.  In the original integrand reduction algorithm
their computation is nevertheless needed, in order to implement the
corresponding subtractions at the integrand level.  Within the Laurent
expansion approach, since the subtraction terms of five-point residues
always vanish in the asymptotic limit, their computation is instead
not needed and can be skipped.

\paragraph{Boxes}
The coefficient $c_0$ of the box contributions can be determined via
4-dimensional quadruple cuts \cite{Britto:2004nc}.  In four dimensions (i.e.\ $\bar
q = q$, $\mu^2=0$) a quadruple cut $D_{i_1}=\cdots=D_{i_4}=0$ has two
solutions, $q_+$ and $q_-$.  The coefficient $c_0$ can be expressed in
terms of these solutions as
\begin{equation}
 c^{(i_1 i_2 i_3 i_4)}_0=\frac{1}{2} \left(\left. \frac{\N(q)}{\prod_{j\neq i_1,i_2,i_3,i_4}D_j}\right|_{q=q_+}+\left.\frac{\N(q)}{\prod_{j\neq i_1,i_2,i_3,i_4}D_j}\right|_{q=q_-}\right).
\end{equation}
The coefficient $c_4$ can be
found by evaluating the integrand on $d$-dimensional quadruple cuts in
the asymptotic limit $\mu^2\rightarrow\infty$~\cite{Badger:2008cm}.  A $d$-dimensional
quadruple cut has an infinite number of solutions which can be
parametrized by the extra-dimensional variable $\mu^2$.  These
solutions become particularly simple in the limit of large $\mu^2$,
namely
\begin{equation}
  q_{\pm}^\nu = a^\nu \pm \sqrt{\mu^2+\beta}\, e_4^\nu\; \overset{\mu^2\rightarrow\infty}{\longrightarrow}\; \pm \sqrt{\mu^2} \, e_4^\nu + \O(1)
\end{equation}
where the vector $a^\nu$ and the constant $\beta$ are fixed by the cut
conditions. The coefficient $c_4$, when non-vanishing, can be found in
this limit as the leading term of the expansion of the integrand
\begin{equation}
    \frac{\N(q,\mu^2)}{\prod_{j\neq i_1,i_2,i_3,i_4}D_j} \; \overset{\mu^2\rightarrow\infty}{\longrightarrow}\; c^{(i_1 i_2 i_3 i_4)}_4\, \mu^4 +\O(\mu^3).
\end{equation}
The other coefficients of the boxes are spurious and their computation
can be avoided.

\paragraph{Triangles}
The residues of the triangle contributions can be determined by
evaluating the integrand on the solutions of $d$-dimensional triple
cuts~\cite{Forde:2007mi}, which can be parametrized by the extra-dimensional variable
$\mu^2$ and another parameter $t$,
\begin{equation}
  \label{eq:triplecutsolutions}
  q_+ = a^\nu+t\, e_3^\nu + \frac{\beta + \mu^2}{2\, t}\, e_4^\nu, \qquad   q_- = a^\nu+ \frac{\beta + \mu^2}{2\, t}\, e_3^\nu+t\, e_4^\nu,
\end{equation}
where the vector $a^\nu$ and the constant $\beta$ are fixed by the cut
conditions $D_{i_1}=D_{i_2}=D_{i_3}=0$.  On these solutions, the
integrand generally receives contributions from the residue of the
triple cut $\Delta_{i_1 i_2 i_3}$, as well as from the boxes and
pentagons which share the three cut denominators.  However, after
taking the asymptotic expansion at large $t$ and dropping the terms
which vanish in this limit, the pentagon contributions vanish, while
the box contributions are constant in $t$ but vanish when taking the
average between the parametrizations $q_+$ and $q_-$ of
Eq.~\eqref{eq:triplecutsolutions}.  More explicitly,
\begin{alignat}{3} \frac{\N(q_\pm,\mu^2)}{\prod_{j\neq i_1,i_2,i_3}D_j} & = \Delta_{i_1 i_2 i_3} + {} & &{}\sum_{j}\frac{\Delta_{i_1 i_2 i_3 j}}{D_j}& & + \sum_{jk}\frac{\Delta_{i_1 i_2 i_3 jk}}{D_jD_k} \nn
  &= \Delta_{i_1 i_2 i_3} + {} & & {} d_1^\pm + d_2^\pm\, \mu^2 & &+ \O(1/t),\quad \qquad \textrm{with }d_i^+ + d_i^- = 0.
\label{eq:triangledecexp}
\end{alignat}
Moreover, even though the integrand is a rational function, in this
asymptotic limit it exhibits the same polynomial behavior as the
expansion of the residue $\Delta_{i_1 i_2 i_3}$,
\begin{align}
          \frac{\N(q_+,\mu^2)}{\prod_{j\neq i_1,i_2,i_3}D_j}  ={}& n_0^++n_7^+\, \mu^2-(n_4+n_9\, \mu^2)\, {t} +n_5\, {t^2} +n_6\, {t^3} +\O(1/{t}) \nonumber \\[0.8ex]
          \frac{\N(q_-,\mu^2)}{\prod_{j\neq i_1,i_2,i_3}D_j} ={}& n_0^-+n_7^-\, \mu^2-(n_1+n_8\, \mu^2)\, {t} +n_2\, {t^2} +n_3\, {t^3} +\O(1/{t})  \pagebreak[1] \label{eq:trianglenumexp} \\[1.2ex]
        \Delta_{i_1 i_2 i_3}(q_+,\mu^2) ={}& c^{(i_1 i_2 i_3)}_0 + c^{(i_1 i_2 i_3)}_7\, \mu^2 - (c^{(i_1 i_2 i_3)}_4+c^{(i_1 i_2 i_3)}_9\, \mu^2)\, {t} \nonumber \\[0.8ex]
        {}&  + c^{(i_1 i_2 i_3)}_5\, {t^2} - c^{(i_1 i_2 i_3)}_6\, {t^3} +\O(1/{t}) \nn
        \Delta_{i_1 i_2 i_3}(q_-,\mu^2) ={}& c^{(i_1 i_2 i_3)}_0 + c^{(i_1 i_2 i_3)}_7\, \mu^2 - (c^{(i_1 i_2 i_3)}_1+c^{(i_1 i_2 i_3)}_8\, \mu^2)\, {t} \nonumber \\[0.8ex]
        {}& + c^{(i_1 i_2 i_3)}_2\, {t^2} - c^{(i_1 i_2 i_3)}_3\, {t^3} +\O(1/{t}) \label{eq:triangledeltaexp}.
\end{align}
By comparison of Eq.s~\eqref{eq:trianglenumexp},
\eqref{eq:triangledeltaexp} and \eqref{eq:triangledecexp} one gets all
the triangle coefficients as
\begin{equation}
  c^{(i_1 i_2 i_3)}_{i}=\frac{1}{2}(n_i^+ + n_i^-) \quad \textrm{for $i=0,7$,} \qquad c^{(i_1 i_2 i_3)}_{i}=n_i\quad \textrm{for $i\neq 0,7$}.
\end{equation}
It is worth to observe that, within the Laurent expansion approach, 
the determination of the 3-point residues does not require any subtraction of higher-point terms.

\paragraph{Bubbles}
The on-shell solutions of a $d$-dimensional double cut
$D_{i_1}=D_{i_2}=0$ can be parametrized as
\begin{align}
  \label{eq:bubblecutsolutions}
  q_+ & = a_0^\nu+x\, a_1^\nu+t\, e_3^\nu + \frac{\beta_0 + \beta_1\, x + \beta_2\, x^2 + \mu^2}{2\, t}\, e_4^\nu, \nn
  q_- & = a_0^\nu+x\, a_1^\nu + \frac{\beta_0 + \beta_1\, x + \beta_2\, x^2 + \mu^2}{2\, t}\, e_3^\nu+t\, e_4^\nu,
\end{align}
in terms of the three free parameters $x$, $t$ and $\mu^2$, while the
vectors $a_i^\nu$ and the constants $\beta_i$ are fixed by the
on-shell conditions.  After evaluating the integrand on these
solutions and taking the asymptotic limit $t\rightarrow\infty$, the
only non-vanishing subtraction terms come from the triangle
contributions,
\begin{align}
  \frac{\N(q_\pm,\mu^2)}{\prod_{j\neq i_1,i_2}D_j} &
  = \Delta_{i_1 i_2} + \sum_{j}\frac{\Delta_{i_1 i_2 j}}{D_j}
  +\sum_{jk}\frac{\Delta_{i_1 i_2 jk}}{D_jD_k} +
  \sum_{jkl}\frac{\Delta_{i_1 i_2 jkl}}{D_jD_kD_l} \nn
  & = \Delta_{i_1 i_2}+\sum_{j}\frac{\Delta_{i_1 i_2 j}}{D_j}+\O(1/t).   \label{eq:bubbledecexp}
\end{align}
The integrand and the subtraction term are rational function, but in
the asymptotic limit they both have the same polynomial behavior as
the residue, namely
\begin{align}
        \frac{\N(q_+,\mu^2)}{\prod_{j\neq i_1,i_2}D_j} ={}& n_0+n_9\, {\mu^2}+n_1\, {x} +n_2\, {x^2} - \big(n_5 + n_8 {x}\big) {t} +n_6\, {t^2} +\O(1/{t}) \nonumber \\[0.8ex]
        \frac{\N(q_-,\mu^2)}{\prod_{j\neq i_1,i_2}D_j} ={}& n_0+n_9\, {\mu^2}+n_1\, {x} +n_2\, {x^2} - \big(n_3 + n_7 {x}\big) {t} +n_4\, {t^2} +\O(1/{t}) \label{eq:bubblenumexp} \pagebreak[1]  \\[1.2ex]
   \frac{\Delta_{i_1 i_2 j}(q_+,\mu^2)}{D_j} ={}&  c_{s,0}^{(j)}+ c_{s,9}^{(j)}\, {\mu^2}+ c_{s,1}^{(j)}\, {x} + c_{s,2}^{(j)}\, {x^2} - \big( c_{s,5}^{(j)} +  c_{s,8}^{(j)} {x}\big) {t} + c_{s,6}^{(j)}\, {t^2}+\O(1/{t}) \nn
   \frac{\Delta_{i_1 i_2 j}(q_-,\mu^2)}{D_j} ={}&  c_{s,0}^{(j)}+ c_{s,9}^{(j)}\, {\mu^2}+ c_{s,1}^{(j)}\, {x} + c_{s,2}^{(j)}\, {x^2} - \big( c_{s,3}^{(j)} +  c_{s,7}^{(j)} {x}\big) {t} + c_{s,4}^{(j)}\, {t^2}+\O(1/{t}) \label{eq:bubblessubexp} \pagebreak[1] \\[1.2ex]
  \Delta_{i_1 i_2}(q_+,\mu^2) ={}& c^{(i_1 i_2)}_0+c^{(i_1 i_2)}_9\, {\mu^2}+c^{(i_1 i_2)}_1\, {x} +c^{(i_1 i_2)}_2 {x^2} - \big(c^{(i_1 i_2)}_5 + c^{(i_1 i_2)}_8 {x}\big)\, {t} \nonumber \\[0.8ex]
  & +c^{(i_1 i_2)}_6\, {t^2}+\O(1/{t}) \nonumber \\[0.8ex]
  \Delta_{i_1 i_2}(q_-,\mu^2)  ={}& c^{(i_1 i_2)}_0+c^{(i_1 i_2)}_9\, {\mu^2}+c^{(i_1 i_2)}_1\, {x} +c^{(i_1 i_2)}_2 {x^2} - \big(c^{(i_1 i_2)}_3 + c^{(i_1 i_2)}_7 {x}\big)\, {t} \nonumber \\[0.8ex] 
   & +c^{(i_1 i_2)}_4\, {t^2}+\O(1/{t}). \label{eq:bubbledeltaexp}
\end{align}
The coefficients $c_{s,i}^{(j)}$ of the Laurent expansion of the
subtractions terms in Eq.s \eqref{eq:bubblessubexp} can be computed
once and for all as parametric functions of the triangles
coefficients.  Therefore, the subtraction of the triangles can be
implemented as corrections to the coefficients of the expansion of the
integrand.  Indeed, by inserting Eq.s~\eqref{eq:bubblenumexp},
\eqref{eq:bubblessubexp} and \eqref{eq:bubbledeltaexp} in
Eq.~\eqref{eq:bubbledecexp} one gets
\begin{equation}
  c^{(i_1i_2)}_i = n_i - \sum_j c_{s,i}^{(j)} \quad \textrm{for $i=0,\ldots,9$}.
  \label{eq:bubcoeffs}
\end{equation}

\paragraph{Tadpoles}
Once the coefficients of the triangles and the bubbles are known, one
can determine the non-spurious coefficient $c_0$ of a tadpole residue
$\Delta_{i_1}$ by evaluating the integrand on the single cut
$D_{i_1}=0$.  One can choose 4-dimensional solutions of the form
\begin{equation}
  q^\nu_+ = -p_{i_1}^\nu + t\, e_3^\nu + \frac{\beta}{2\, t}\, e_4^\nu
\end{equation}
parametrized by the free variable $t$, while the constant $\beta$ is
fixed by the cut conditions.  In the asymptotic limit $t\rightarrow
\infty$, only bubbles and triangles coefficients are non-vanishing,
\begin{align} \frac{\N(q_+)}{\prod_{j\neq i_1}D_j} & = \Delta_{i_1} 
+ \sum_{j}\frac{\Delta_{i_1  j}}{D_j}+ \sum_{jk}\frac{\Delta_{i_1 jk}}{D_j D_k}+ \sum_{jkl}\frac{\Delta_{i_1 jkl}}{D_jD_k D_l} \nn
&= \Delta_{i_1}+ \sum_{j}\frac{\Delta_{i_1  j}}{D_j}+ \sum_{jk}\frac{\Delta_{i_1 jk}}{D_j D_k} + \O(1/t).
\label{eq:tadpoledecexp}
\end{align}
Similarly to the case of the bubbles, in this limit the integrand and
the subtraction terms exhibit the same polynomial behavior as the
residue, i.e.\
\begin{align}
  \frac{\N(q_+)}{\prod_{j\neq i_1}D_j} & = n_0 - n_4\, t + \O(1/t) \\[0.5ex]
  \frac{\Delta_{i_1 j}(q_+)}{D_j} & = c_{s_2,0}^{(j)} - c_{s_2,4}^{(j)}\, t + \O(1/t) \\[0.5ex]
  \frac{\Delta_{i_1 j k}(q_+)}{D_j D_k} & = c_{s_3,0}^{(jk)} - c_{s_3,4}^{(jk)}\, t + \O(1/t) \\[0.5ex]
  \Delta_{i_1}(q_+) & = c^{(i_1)}_0 - c^{(i_1)}_4\, t.
\end{align}
Putting everything together, we can write the coefficient of the
tadpole integral as the corresponding one in the expansion of the
integrand, corrected by coefficient-level subtractions
\begin{equation}
  c^{(i_1)}_0 = n_0 - \sum_j c_{s_2,0}^{(j)}  - \sum_{jk} c_{s_3,0}^{(jk)}.
  \label{eq:tadcoeffs}
\end{equation}
Once again, we observe that the subtraction terms $c_{s_2,0}^{(j)}$
and $c_{s_3,0}^{(jk)}$, coming from bubbles and triangles
contributions respectively, are known parametric functions of the
coefficients of the corresponding higher-point residues.

\subsection{The {\sc C++} library {\sc Ninja}}
\label{sec:ninjalibrary}
The {\sc C++} library {\sc Ninja}~\cite{Peraro:2013oja, Peraro:2014cba} provides a semi-numerical
implementation of the Laurent expansion method described above.  Since
the integrand is a rational function of the loop variables, its
Laurent expansion is performed via a simplified polynomial division
algorithm between the expansion of the numerator $\N$ and the uncut
denominators.

The inputs of the reduction algorithm implemented in {\sc Ninja} are
the external momenta $p_i$ and the masses $m_i$ of the loop
denominators defined in Eq.~\eqref{eq:loopdenom}, and the numerator
$\N(q,\mu^2)$ of the integrand cast in four different forms.  
\begin{itemize}
\item The first form provides a simple evaluation of the numerator as a function
of $q$ and $\mu^2$, which is used in order to compute the coefficient
$c_0$ of the boxes.
It can also be used in order to compute the spurious coefficients of
the pentagons via penta-cuts, and all the ones of the boxes when the
expansion in $\mu^2$ is not provided.  
\end{itemize}
The other three forms of the
numerator yield instead the leading terms of a parametric expansion of
the integrand.

\begin{itemize}
\item The first expansion is the one used in order to
obtain the coefficient $c_4$ of the boxes.  When the rank $r$ is equal
to the number $n$ of external legs of the diagram, this method returns
the coefficient of $t^n$ obtained by substituting in the numerator
$\N(q,\mu^2)$
\begin{equation}
  q^\nu\rightarrow t\, v^\nu,\qquad \mu^2 \rightarrow t^2\, v^2
\end{equation}
as a function of a generic vector $v$, which is computed by {\sc
  Ninja} and is determined by the quadruple-cut constraints.

\item The second expansion is used in order to compute triangles and tadpole
coefficients.  In this case it returns coefficients of the terms
$t^j\mu^{2k}$ for $j=r,r-1,\ldots,n-3$, obtained from
$\N(q,\mu^2)$ with the substitutions
\begin{equation}
  \label{eq:ninjat3exp}
  q^\nu\rightarrow v_0^\nu + t\, v_3^\nu + \frac{\beta+\mu^2}{t}\, v_4^\nu, \qquad  v_3^2=v_4^2=0,
\end{equation}
as a function of the generic momenta $v_i^\nu$ and the constant
$\beta_0$.  Since in a renormalizable theory $r \leq n$, and by
definition of rank we have $j+2k\leq r$, at most 6 terms can be
non-vanishing in the specified range of $j$.  For effective theories
with $r \le n+1$, one can have instead up to 9 non-vanishing
polynomial terms.  In each call of the numerator, {\sc Ninja}
specifies the lowest power of $t$ which is needed in the evaluation,
avoiding thus the computation of unnecessary coefficients of the
expansion.

\item 
The third and last expansion is needed for the computation of the
2-point residues and returns the coefficients of the terms
$t^j\mu^{2k}x^l$ for $j=r,r-1,\ldots,n-2$, obtained from $\N(q,\mu^2)$
with the substitutions
\begin{equation}
  \label{eq:ninjat2exp}
  q^\nu\rightarrow v_1^\nu + x\, v_2^\nu + t\, v_3^\nu + \frac{\beta_0+\beta_1\, x+\beta_2\, x^2+\mu^2}{t}\, v_4^\nu, \qquad v_2\cdot v_3=v_2\cdot v_4=v_3^2=v_4^2=0,
\end{equation}
as a function of the cut-dependent momenta $v_i^\nu$ and constants
$\beta_i$.  In a renormalizable theory one can have at most 7
non-vanishing terms in this range of $j$, while for $r\leq n+1$ one
can have 13 non-vanishing terms.  As in the previous case, in each
call of the numerator, {\sc Ninja} specifies the lowest power of $t$
which is needed.  It is worth to notice that the expansion in
Eq.~\eqref{eq:ninjat2exp} can be obtained from the previous one in
Eq.~\eqref{eq:ninjat3exp} with the substitutions
\begin{equation*}
  v_0^\nu \rightarrow v_1^\nu + x\, v_2^\nu, \qquad \beta\rightarrow \beta_0+\beta_1\, x+\beta_2\, x^2, \qquad v_2\cdot v_3=v_2\cdot v_4=0.
\end{equation*}

\end{itemize}
All these expansions can be easily and quickly obtained from the
knowledge of the analytic dependence of the loop momentum on $q$ and
$\mu^2$.  For every phase-space point, {\sc Ninja} computes the
parametric solutions of all the multiple cuts, performs the Laurent
expansion of the integrand via a simplified polynomial division
between the expansion of the numerator and the set of the uncut
denominators, and implements the subtractions at the coefficient level
appearing in Eqs.~\eqref{eq:bubcoeffs} and~\eqref{eq:tadcoeffs}.
Finally, the obtained non-spurious coefficients are multiplied by the
corresponding Master Integrals in order to get the integrated result
as in Eq.~\eqref{eq:integraldecomposition}.

The routines which compute the Master Integrals are called by {\sc
  Ninja} via a generic interface which allows to use any integral
library implementing it, with the possibility of switching between
different libraries at run-time.  By default, a {\sc C++} wrapper of
the {\sc OneLoop} integral library~\cite{vanHameren:2010cp,vanHameren:2009dr} is used.  This wrapper
caches every computed integral allowing constant time lookups of their
values from their arguments.  An interface with the {\sc LoopTools}~\cite{Hahn:1998yk, Hahn:2010zi}
library is available as well.

The {\sc Ninja} library can also be used in order to compute integrals
from effective and non-renomalizable theories where the rank $r$ of
the numerator can exceed the number of legs by one unit.  An
example of this application, given in Section \ref{sec:Applications},
is Higgs boson production plus three jets in gluon fusion, in the
effective theory defined by the infinite top-mass limit.

%The simplified fit of the coefficients and the subtractions at
%coefficient level make the algorithm implemented in {\sc Ninja}
%significantly lighter, faster and more stable than the original.

%%%%%%%%%%%%%%%%%%%%%%%%%%%%%%%%%%%%%%%%
\section{Interfacing \ninja\ with \gosam} \label{sec:Ninja}

The library \Ninja{} has been interfaced with the automatic generator
of one-loop amplitudes {\sc GoSam}. The latter provides \Ninja{} with analytic
expressions for the integrands of one-loop Feynman diagrams for
generic processes within the Standard Model and also for Beyond
Standard Model theories.

{\gosam} combines automated diagram generation and algebraic manipulation~\cite{Nogueira:1991ex, Vermaseren:2000nd, Reiter:2009ts, Cullen:2010jv} with integrand-reduction
techniques~\cite{Ossola:2006us, Ossola:2007bb, Ellis:2007br, Ossola:2008xq,  Mastrolia:2008jb,Mastrolia:2012bu}.
Amplitudes are generated via Feynman diagrams, using \QGRAF~\cite{Nogueira:1991ex}, \FORM~\cite{Vermaseren:2000nd}, \SPINNEY~\cite{Cullen:2010jv} and \HAGGIES~\cite{Reiter:2009ts}.

After the generation of all contributing diagrams, the virtual corrections are evaluated using the $d$-dimensional integrand-level reduction method,
as implemented in the library \SAMURAI~\cite{Mastrolia:2010nb}, which allows for the combined
determination of both cut-constructible and rational terms at once.
As an alternative, the tensorial decomposition provided by
{\Golem}~\cite{Binoth:2008uq,Heinrich:2010ax,Cullen:2011kv}  is also
available. Such reduction, which is numerically stable but more time
consuming, is employed as a rescue system.
After the reduction, all relevant master integrals can be computed by
means of {\Golem}~\cite{Cullen:2011kv},
{\QCDLoop}~\cite{vanOldenborgh:1990yc, Ellis:2007qk}, or {\OneLoop}~\cite{vanHameren:2010cp}.

The possibility to deal with higher-rank one-loop integrals, where powers of loop
momenta in the numerator exceed the number of denominators, is implemented in all three 
reduction programs \samurai{}~\cite{Mastrolia:2012bu,vanDeurzen:2013pja}, \Ninja{} and
\GOLEMVC{}~\cite{Guillet:2013msa}. Higher rank integrals can appear when computing one-loop integrals in effective-field theories, e.g. for calculations involving the effective gluon-gluon-Higgs
vertex~\cite{vanDeurzen:2013rv,Cullen:2013saa}, or when dealing with spin-2 particles~\cite{Greiner:2013gca}. 

In order to embed \Ninja{} into the \Gosam{} framework,
the algebraic manipulation of the integrands was adapted to generate
the expansions needed by \Ninja{} and described in Section~\ref{sec:ninjalibrary}. 
The numerator, in all its forms, is then optimized for fast numerical
evaluation by exploiting the new features of {\sc Form 4}~\cite{Kuipers:2012rf,Kuipers:2013pba}, 
and written in a {\sc Fortran90} source file 
which is compiled. At running time, these expressions are used as input for \Ninja{}.

The {\sc Fortran90} module of the interface between \Ninja{} and
\Gosam{} defines subroutines which allow to control some of the settings
of \Ninja{} directly from settings of the code that generated the virtual part of the amplitudes. Upon importing
the module, 
%$\tt{ninjago\_module}$, 
the user can change the integral library used by \Ninja{} 
%by calling
%~\\
%$$\tt{call\;ninja\_set\_integral\_library(libflag)},$$
%~\\
%where the argument $\tt{libflag}$ can either be $\tt{NINJA\_ONELOOP}$,
choosing between the use of \OneLoop{}~\cite{vanHameren:2010cp} and 
%$\tt{NINJA\_LOOPTOOLS}$, for 
the {\sc LoopTools}~\cite{Hahn:1998yk, Hahn:2010zi}.  

For debugging purposes, one can also ask
\Ninja{} to perform some internal test or print some information about
the ongoing computation. 
%The internal tests of \Ninja{} can be
%activated and changed with the call
%~\\
%$$\tt{call\;ninja\_set\_test(val)}.$$
%~\\
%Different values for the argument $\tt{val}$ 
This option allows to choose among different internal tests, namely the global $\N=\N$ test,  the local $\N=\N$ tests on different cuts, or a combination of both. These tests have been described in~\cite{Mastrolia:2010nb}.
%Examples of values for the argument $\tt{val}$ are
%\begin{itemize}
%\item[1:]{$\tt{NINJA\_TEST\_NONE}$: no test (default value),}
%\item[2:]{$\tt{NINJA\_TEST\_ALL}$: all tests,}
%\item[3:]{$\tt{NINJA\_TEST\_GLOBAL}$: global $\N=\N$ tests,}
%\item[4:]{$\tt{NINJA\_TEST\_LOCAL}$: all local $\N=\N$ tests,}
%\item[5:]{$\tt{NINJA\_TEST\_LOCAL\_}\it{k}$: local $\N=\N$ tests on $k$-ple cuts ($k=1,\ldots,4$),}
%\end{itemize}
%and different flags can also be combined with the $\tt{IOR}$ function.
%If a test fails, the return status flag will be set to
%$\tt{NINJA\_TEST\_FAILED}$, otherwise it will be set to
%$\tt{NINJA\_SUCCESS}$.  The subroutine
%~\\
%$$\tt{call\;ninja\_set\_test\_tolerance(val)}$$
%~\\
%sets the maximum tolerance on the relative error of a successful
%test.  
The verbosity of the debug output can be adjusted to control the  
%with
%~\\
%$$\tt{call\;ninja\_set\_verbosity(val)}.$$
%~\\
%Different values for the argument $\tt{val}$ 
%to choose the 
amount of details printed out in the output file, for example the final results for the finite part and the poles of the diagram, the values of the coefficients that are computed in the reduction, the values of the corresponding Master Integrals, and the results of the various internal tests. 
%Examples of values for the argument $\tt{val}$ are
%\begin{itemize}
%\item[1:]{$\tt{NINJA\_OUTPUT\_NONE}$: nothing is printed (default value),}
%\item[2:]{$\tt{NINJA\_OUTPUT\_ALL}$: everything is printed,}
%\item[3:]{$\tt{NINJA\_OUTPUT\_TEST\_GLOBAL}$: the result of the global test is printed,}
%\item[4:]{$\tt{NINJA\_OUTPUT\_TEST\_LOCAL}$: the result of the local tests is printed,}
%\item[5:]{$\tt{NINJA\_OUTPUT\_TEST\_LOCAL\_}\it{k}$: the result of the local tests on $k$-ple cuts is printed,}
%\item[6:]{$\tt{NINJA\_OUTPUT\_TESTS}$: the result of all the performed tests are printed,}
%\item[7:]{$\tt{NINJA\_OUTPUT\_COEFFS}$: all the computed coefficients are printed,}
%\item[8:]{$\tt{NINJA\_OUTPUT\_C}\it{k}$: the computed coefficients of $k$-ple cuts are printed ($k=1,\ldots,5$),}
%\item[9:]{$\tt{NINJA\_OUTPUT\_RESULT}$: the final result is printed,}
%\item[10:]{$\tt{NINJA\_OUTPUT\_INTEGRALS}$: the values of the computed Master Integrals are printed.}
%\end{itemize}
%Once again, different flags can be combined with the $\tt{IOR}$
%function.

\section{Precision tests} \label{sec:Precision}

Within the context of numerical and semi-numerical techniques, the problem of estimating correctly the precision of the results is of primary importance. In particular, when performing the phase space integration
of the virtual contribution, it is important to assess in real time, for each phase space point, the level of precision of the corresponding one-loop matrix element. 

Whenever a phase space point is found in which the quality of the result falls below a certain threshold, either the point is discarded or the evaluation of the amplitude is repeated by means of a safer, albeit less efficient procedure.
Examples of such a method involve the use of higher precision routines, or in the case of \gosam{} the use of traditional tensorial reconstruction of the amplitude, provided by \Golem{}.

Various techniques for detecting points with low precision have been implemented within the different automated tools for the evaluation of one-loop virtual corrections.

A standard method which is widely employed is based on the comparison between the numerical values of the poles with their known analytic results dictated by the universal behavior of the infrared singularities. While this method is quite reliable, not all integrals which appear in the reconstruction of the amplitude give a contribution to the double and single poles. This often results in an overestimate of the precision, which might lead to keep phase space points whose finite part is less precise than what is predicted by the poles.

A different technique, which we refer to as {\it scaling test}~\cite{Badger:2010nx}, is based on the properties of scaling of scattering amplitudes when all physical scales (momenta, renormalization scale, masses) are rescaled by a common multiplicative factor $x$. As shown in~\cite{Badger:2010nx}, this method provides a very good correlation between the estimated precision, and the actual precision of the finite parts.

Additional methods have been proposed, within the context of integrand-reduction approaches, which target the relations between the coefficients before integration, namely the reconstructed algebraic expressions for the numerator function before integration.  This method, labeled $\N=\N$ test~\cite{Ossola:2007ax, Mastrolia:2010nb}, can be applied to the full amplitude (global   $\N=\N$ test) or individually within each residue of individual cuts (local $\N=\N$ test). The drawback of this technique comes from the fact that the test is applied at the level of individual diagrams, rather than on the final result summed over all diagrams, making the construction of a rescue system quite cumbersome. 

For the precision analysis contained in this paper, % in addition to employing the techniques described above, 
we present a new simple and efficient method for the estimation of the number of digits of precision in the results, which we call {\it rotation test}. This new method exploits the invariance of the scattering amplitudes under an azimuthal rotation about the beam axis, namely the direction of the initial colliding particles. 

Such a rotation, which does not affect the initial states, changes the momenta of all final particles without changing their relative position, thus reconstructing a theoretically identical process. However, the change in the values of all final state external momenta is responsible for different bases for the parametrization of the residues within the integrand reconstruction, different coefficients in front of the master integrals, as well as different numerical values when the master integrals are computed.
%In order to validate the {\emph rotation test}, we compared it with other techniques and the results, when compared to the pole test and the rescaling test, are indeed very encouraging.
% In addition, 
We tested that the choice of the angle of rotation does not affect the estimate, as long as this angle is not too small.
% As a final test, we compared the prediction of the rotation test with the absolute errors, obtained by employing quadrupole precision routines.

In order to study the correlation of the error estimated by the rotation test and the exact error, we follow the strategy of Ref.~\cite{Badger:2010nx}. In particular, we generated $10^4$ points for the process $u \bar d \to W b \bar b g$ with massive bottom quarks, both in quadrupole and standard double precision, which we label with $A_{quad}$ and $A$ respectively, as well as the same points in double precision after performing a rotation, called $A_{rot}$.

We define the exact error $ \delta_{ex}$ as 
\beq \label{eq:exd}
\delta_{ex} = \left | \frac{ A_{quad} - A }{ A_{quad}} \right |\, ,
\eeq
and the estimated error $ \delta_{rot}$  as  
\beq  \label{eq:errd} \delta_{rot} =  2 \left  |\frac{ A_{rot} - A }{ A_{rot} + A} \right  |\, . \eeq

\begin{figure*}[ht]
\centering
\includegraphics{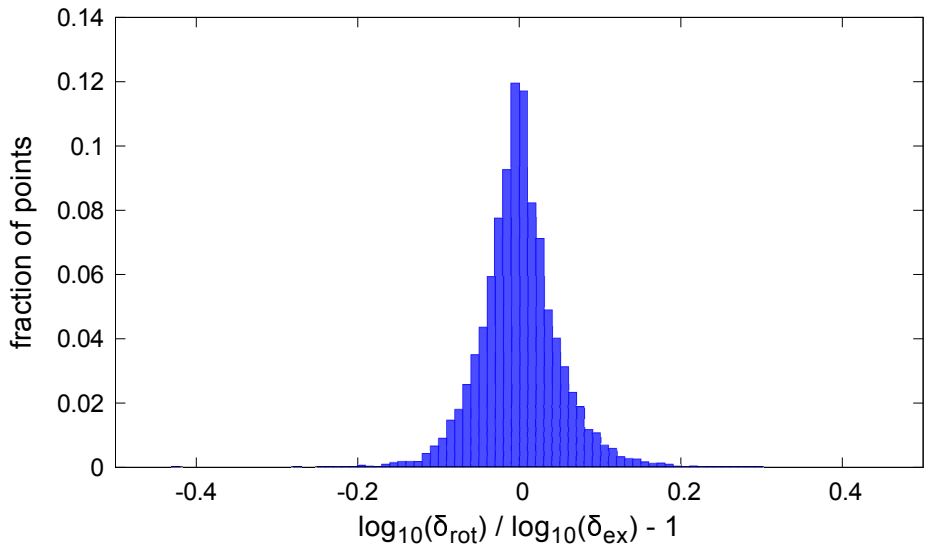}
\caption{Correlation plot based on $10^4$ points for the process $u \bar d \to W b \bar b g$ with massive bottom quarks }
\label{corr}
\end{figure*}

In Fig.~\ref{corr}, we plot the distribution of the quantity 
\beq {\cal C} = \frac{ \log_{10} (\delta_{rot})}{\log_{10} (\delta_{ex})}-1 \, , \eeq
evaluated for each phase space point.
In the ideal case of a perfect correlation between the exact error, as estimated by the quadrupole precision result, and the error estimated by the less time-consuming rotation test, the value of ${\cal C} $ would be close to zero, while the spread of the distribution can provide a picture of the degree of correlation.
Moreover, we observe a similar behavior for the rotation and the scaling tests.

In the following, we will employ the rotation test as the standard method to estimate the precision of the finite part of each renormalized virtual matrix elements. If we call $\delta_{0}$ the error at any given phase space point and calculate it according to Eq.~\eqref{eq:errd}, we can define the precision of the finite part as  $P_0 = \log_{10} (\delta_{0})$.
Concerning the precision of the double and single poles, $P_{-2} = \log_{10} (\delta_{-2})$ and $P_{-1}= \log_{10} (\delta_{-1})$, we employ the fact that the values of the pole coefficients, after renormalization, are solely due to infrared (IR) divergencies, whose expressions are well known~\cite{Catani:2000ef}. $\delta_{-2}$ and $\delta_{-1}$ are defined using formula in Eq.~\eqref{eq:exd}, in which the exact values are provided by the reconstructed IR poles, which is automatically evaluated by \gosam.

\begin{figure}[t]
\begin{center}
\includegraphics{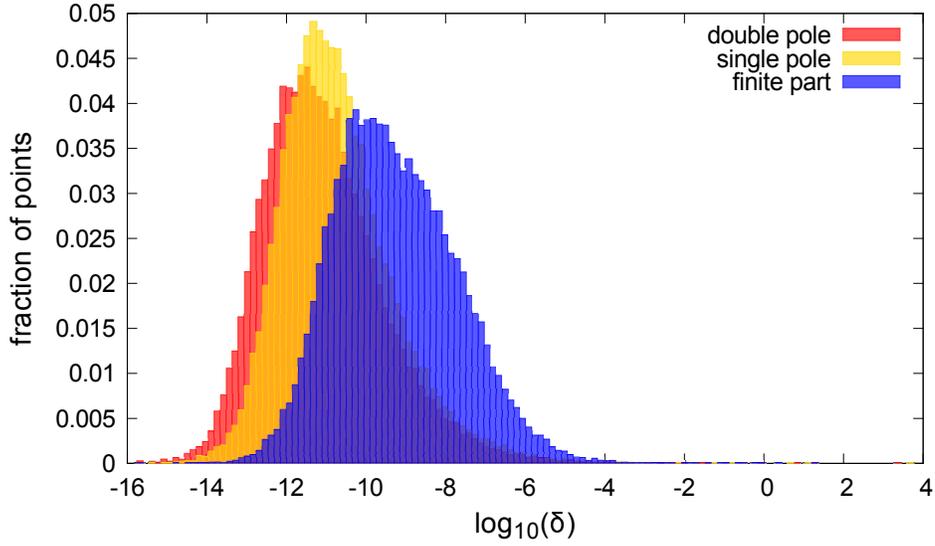} 
\caption{Precision Plot for $g g \to t \bar t H g$: the distributions are obtained using  $5 \cdot 10^4$ randomly distributed phase space points. }
\label{precA}
\end{center}
\end{figure}

\begin{figure*}[t]
\centering
\includegraphics{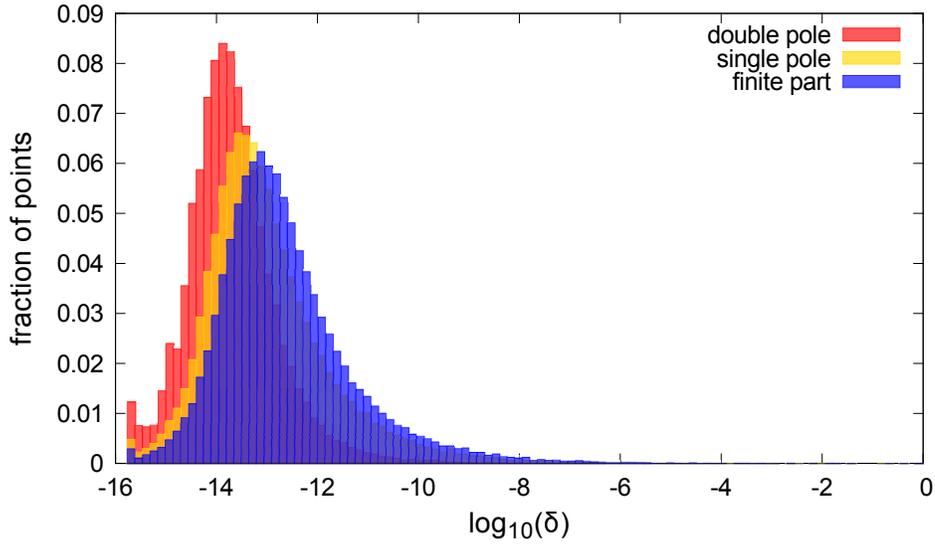}
\caption{Precision plot for $u \bar u \to H u \bar u g g $ in VBF: the distributions are obtained using  $10^5$ randomly distributed phase space points.}
\label{precB}
\end{figure*}

In order to assess the level of precision of the results obtained with  \ninja\ within \gosam{}, 
in Figs.~\ref{precA} and~\ref{precB}, we plot the distributions of $P_{-2}$ (precision of the double pole), $P_{-1}$ (single pole) and  $P_{0}$ (finite part) for two challenging virtual amplitudes with massive internal and external particles, namely  $g g \to t \bar t H g$ ($t \bar t H j$) and $u \bar u \to H u \bar u g g $ ($Hjjjj$) in VBF. 
By selecting an upper bound on the value of $P_{0}$, we can set a {\it rejection criterium} for phase space points in which the quality of the calculated scattering amplitudes falls below the requested precision.  This also allows to estimate the percentage of points which would be discarded (or redirected to the  rescue system). This value, as expected by analyzing the shape of the various distributions, is strongly process dependent and should be selected according to the particular phenomenological analysis at hand.
As a benchmark value, in Ref.~\cite{Badger:2010nx}, the threshold for rejection was set to $P_{0}=-3$.
In a similar fashion, in Table~\ref{tab:bad}, we provide the percentages of {\it bad points}, which are points whose precision falls below the threshold, for increasing values of the rejection threshold. 

The two plots %in Figs.~\ref{precA} and~\ref{precB} 
are built using a set of  $5 \cdot 10^4$  and $1\cdot10^5$ phase space points, respectively for $g g \to t \bar t H g$ and 
$u \bar u \to H u \bar u g g $ (VBF). No cuts have been introduced in the selection of the points, which are randomly distributed over the whole available phase space for the outgoing particles, and are generated using {\sc rambo}.

\begin{table}[h]
\begin{center}
{\small
  \begin{tabular}{ c  c  c }
    \hline
    \hline
  $P_0$   &  $u \bar u \to H u \bar u g g $  &  $g g \to t \bar t H g$   \\
    \hline
  $-3$  & 0.02\% &   0.06\% \\
  $ -4 $    &     0.04\%  &  0.16\%  \\
     $-5$     &  0.08\% &   0.56\% \\
\hline
    \hline
\end{tabular}
}
\caption{Percentage of {bad points} as a function of the rejection threshold  $P_0$.
%, for the distributions in Figs..~\ref{precA} and~\ref{precB}.
} 
\label{tab:bad}
\end{center}
%}
\end{table}

The use of the novel algorithm implemented in \ninja\  yields significant improvements both in the accuracy of results and in reduction of the computational time, due to a more efficient reduction and less frequent calls to the rescue system.

These features make \gosam{}+\ninja\ an extremely competitive framework for massive, as well as massless, calculations.
The new library has been recently used  in the evaluation of NLO QCD corrections to $p p \to t {\bar t} H j$~\cite{vanDeurzen:2013xla}.

\section{Applications to Massive Amplitudes} \label{sec:Applications}
In order to demonstrate the performances of the new reduction algorithm, we
apply {\Gosam}+{\Ninja} to a collection of processes involving 
{\it six},
{\it seven} and 
{\it eight} external particles. We choose processes where massive
particles appear in the products of the reactions or run in the loop. 
We list them in
Table~\ref{tab:summary}, and give the details of their calculations in
the following subsections: for each process we provide results for a phase space point and a detailed list of 
the input parameters employed. 
While some of the considered processes have already been studied in the
literature, the virtual NLO QCD contributions to 
$pp \to W b {\bar b} + n$ jets ($n=1,2$), 
$pp \to Z b {\bar b} j$, $pp \to Z t {\bar t} j$,
$pp \to VVV  j$ (with $V=W,Z$),  $pp \to ZZZZ$, 
and 
$pp \to H + n$ jets ($n=4,5$) in VBF are presented in this paper for the first time.
When occurring in the final state, the {\it bottom} quark is treated as massive.

For calculation which were already performed with previous versions of the {\gosam}  framework,
we observe that the new reduction technique yields a significant net gain both in computing time and in accuracy.
A paradigmatic example is represented by $gg \to Hggg$, whose
evaluation per ps-point required approximately 20 seconds, as reported in~\cite{Cullen:2013saa}, while now can be computed in less than 10 seconds.

%In the following subsections we present numerical results for the renormalized virtual NLO QCD 
%contributions to several processes, which are summarized in Table~\ref{tab:summary}. 
In the following, for each of the considered scattering amplitudes, we provide a benchmark phase space point
for the most involved subprocesses and, when possible, a comparison with results available in the literature.
% All results are computed using dimensional reduction.  
The  coefficients $a_i$ are which appear in the various tables are defined as follows:
\bea
\frac{a_{-2}}{\epsilon^2} +  \frac{a_{-1}}{\epsilon} + a_0   \equiv \frac{  2 \mathfrak{Re} \left  \{ \mathcal{M}^{\mbox{\tiny tree-level} \ast } \mathcal{M}^{\mbox{\tiny one-loop}  }   \right  \}  }{
( \alpha_s / 2 \pi) \left |  \mathcal{M}^{\mbox{\tiny tree-level}} \right |^2 \nonumber
 }  \, ,
\label{Eq:AI}
\eea
where the finite part $a_0$ is computed in the dimensional reduction scheme if not stated otherwise.
The reconstruction  of the renormalized pole can be checked against the value of $a_{-1}$ and $a_{-2}$ obtained by the universal singular behavior of the dimensionally regularized  one-loop
amplitudes~\cite{Catani:2000ef}, while the precision of the finite parts is estimated by re-evaluating the amplitudes for a set of momenta rotated by an arbitrary angle about the axis of collision, as
described in Section~\ref{sec:Precision}. The accuracy of the results obtained with  {\Gosam}+{\Ninja} is indicated by the underlined digits. 

\begin{table}  
\begin{center}
        \begin{tabular}[ht]{|l|l|r|r|}
          \hline
          \multicolumn{4}{|c|}{Benchmarks: {\sc GoSam} + {\sc Ninja}} \\ 
          \hline
          \hline
          \multicolumn{2}{|l|}{Process} & \# NLO diagrams &  ms/event \\
          \hline
          \hline
          \multirow{1}{*}{$ W + 3\, j $} 
          & ${d}{\bar{u}}\rightarrow{\bar{\nu}_e}{e^-}{g}{g}{g}$ & 1 411 &  226 \\
          \hline
          \hline
          \multirow{1}{*}{$ Z + 3\, j $} 
          & ${d}{\bar{d}}\rightarrow{e^+}{e^-}{g}{g}{g}$ & 2 928 & 1 911 \\
          \hline
          \hline
           \multirow{1}{*}{$ Z\, Z\, Z+ 1\, j $} 
          & ${u}{\bar{u}}\rightarrow{Z}{Z}{Z}{g}$ & 915 & *12 000  \\
          \hline
          \hline
     \multirow{1}{*}{$ {W}\,{W}\,{Z}+ 1\, j $} 
          &  ${u}{\bar{u}}\rightarrow{W^+}{W^-}{Z}{g}$ & 779 & *7 050  \\
          \hline
          \hline
           \multirow{1}{*}{$ {W}\,{Z}\,{Z}+ 1\, j $} 
          & ${u}{\bar{d}}\rightarrow{W^+}{Z}{Z}{g}$ & 756 & *3 300  \\
          \hline
          \hline
           \multirow{1}{*}{${W}\,{W}\,{W}+ 1\, j $} 
          & ${u}{\bar{d}}\rightarrow{W^+}{W^-}{W^+}{g}$& 569 & *1 800 \\
           \hline
          \hline
           \multirow{1}{*}{$ Z\, Z\, Z\, Z $} 
          & ${u}\,{\bar{u}}\rightarrow{Z}\,{Z}\,{Z}\,{Z}$ & 408 & *1 070 \\\hline
          \hline
           \multirow{1}{*}{$ W\, W\, W\, W $} 
          & ${u}{\bar{u}}\rightarrow{W^+}{W^-}{W^+}{W^-}$ & 496 & *1 350 \\
          \hline
          \hline      
          \multirow{2}{*}{$ t \bar t b \bar b \, (m_b\neq 0)$} 
          & $\vphantom{e^+}d\bar d\rightarrow t\bar{t}b\bar b$ & 275 & 178 \\
          \cline{2-4}
          & $\vphantom{e^+}gg\rightarrow t\bar{t}b\bar b$ & 1 530 &  5 685 \\
          \hline
          \hline        
        \multirow{1}{*}{$ t \bar t + 2\, j$}
           & $\vphantom{e^+}gg\rightarrow t\bar t gg$ & 4 700 &  13 827 \\
          \hline
          \hline
          \multirow{1}{*}{$\vphantom{e^+}  Z\, b\, \bar b + 1\, j\, (m_b\neq 0) $} 
          &  $\vphantom{e^+} d{u}{g}\rightarrow{u}{e^+}{e^-}{b}{\bar{b}}$ & 708 & *1 070 \\
          \hline
          \hline 
          \multirow{1}{*}{$\vphantom{e^+}  W\, b\, \bar b + 1\, j\, (m_b\neq 0) $} 
          & $\vphantom{e^+} u\bar d\rightarrow e^+\nu_eb\bar{b}g$ & 312 &  67 \\
          \hline
          \hline 
          \multirow{3}{*}{$\vphantom{e^+}  W\,  b\, \bar b + 2\, j\, (m_b\neq 0) $} 
            & ${u}{\bar{d}}\rightarrow{e^+}{\nu_e}{b}{\bar{b}}{s}{\bar{s}}$ & 648 & 181 \\
           \cline{2-4}
             & ${u}{\bar{d}}\rightarrow{e^+}{\nu_e}{b}{\bar{b}}{d}{\bar{d}}$ &  1 220  & 895\\
             \cline{2-4}
             & ${u}{\bar{d}}\rightarrow{e^+}{\nu_e}{b}{\bar{b}}{g}{g}$ & 3 923 & 5387 \\
          \hline
          \hline 
           \multirow{2}{*}{$\vphantom{e^+}  W\,  W\, b\, \bar b \, (m_b\neq 0) $} 
          &  ${d}{\bar{d}}\rightarrow{\nu_e}{e^+}{\bar{\nu}_\mu}{\mu^-}{b}{\bar{b}}$ & 292 & 115  \\
           \cline{2-4}
            & ${g}{g}\rightarrow{\nu_e}{e^+}{\bar{\nu}_\mu}{\mu^-}{b}{\bar{b}}$ & 1 068 & *5 300 \\
          \hline
          \hline 
             \multirow{1}{*}{$\vphantom{e^+}  W\,  W\, b\, \bar b + 1\, j\, (m_b = 0) $} 
          &  ${u}{\bar{u}}\rightarrow{\nu_e}{e^+}{\bar{\nu}_\mu}{\mu^-}{b}{\bar{b}}{g}$ & 3 612 & *2 000\\
%           \cline{2-4}
  %          & & & \
          \hline
          \hline 
          \multirow{1}{*}{$ H + 3\, j$ in GF}
          & $\vphantom{e^+} g g  \rightarrow H g g g$ & 9 325 &  8 961 \\
            \hline
          \hline        
             \multirow{2}{*}{$\vphantom{e^+}  {t}\, {\bar{t}}\, Z\, + 1\, j $} 
          & ${u}{\bar{u}}\rightarrow{t}{\bar{t}}{e^+}{e^-}{g}$  & 1408  &  1 220 \\
           \cline{2-4}
            & ${g}{g}\rightarrow{t}{\bar{t}}{e^+}{e^-}{g}$ &  4230 &  19 560\\
          \hline
          \hline
                    \multirow{1}{*}{$ t\, \bar t\, H + 1\, j$} 
          %& $\vphantom{e^+}q \bar q \rightarrow t \bar t H g$ & 320& 80 \\
          %\cline{2-4}
          & $\vphantom{e^+}g g \rightarrow t \bar t H g$ & 1 517 & 1 505 \\
          \hline
          \hline         
          $ H + 3\, j $ in VBF
          & $\vphantom{e^+} {u}{\bar{u}}\rightarrow H{g}u\bar u$ & 432 &  101 \\
          \hline
          \hline
          $ H + 4\, j $ in VBF
          & $\vphantom{e^+} {u}{\bar{u}}\rightarrow H{g}{g}u\bar u$ & 1 176 &  669 \\
          \hline
          \hline
          $ H + 5\, j $ in VBF
          & $\vphantom{e^+} {u}{\bar{u}}\rightarrow H{g}{g}{g}u\bar u$ & 15 036 &  29 200 \\
          \hline
  \end{tabular}
  \end{center}
\caption{A summary of results obtained with  {\Gosam}+{\Ninja}. Timings refer to full color- and helicity-summed amplitudes, using an Intel Core i7 CPU @ 3.40GHz, compiled with {\tt ifort}. The timings indicated with an (*) are obtained with an Intel(R) Xeon(R) CPU E5-2650 0 @ 2.00GHz, compiled with {\tt gfortran}. }
\label{tab:summary}
\end{table}

\newpage

\begin*

\subsection{$p\,p\rightarrow W+3$ jets}
\paragraph*{Partonic process: ${d}{\bar{u}}\rightarrow{\bar{\nu}_e}{e^-}{g}{g}{g}$}
~\\
The finite part for this process is given in the conventional dimensional regularization (CDR) scheme and was compared to the new version of \njet~\cite{Badger:2012pg}. We also find agreement in the
phase space point given by the
\blackhat{} Collaboration~\cite{Berger:2009ep}.
\begin{table}[ht]
%\TABLE{
%\centering
\begin{center}
{\footnotesize
\begin{tabular}{c c c c c}
\hline 
\hline
\textsc{particle} & $E$& $p_x$ & $p_y$ & $p_z$ \\ 
\hline
$p_1$ & 500.0000000000000000 & 0.0000000000000000 & 0.0000000000000000 & 500.0000000000000000 \\
$p_2$ & 500.0000000000000000 & 0.0000000000000000 & 0.0000000000000000 & -500.0000000000000000 \\
$p_3$ & 414.1300683745429865 & 232.1455649459389861 & 332.7544367808189918 & -82.9857518524426041 \\
$p_4$ & 91.8751521026383955 & -43.3570226791010995 & -24.0058236140056991 & 77.3623460793434958 \\
$p_5$ & 86.3540681437814044 & -15.2133893202618005 & 37.6335512949163018 & -76.2187226821854011 \\
$p_6$ & 280.1181818093759830 & -83.1261116505822031 & -263.2038567586509998 & 47.7490851160265990 \\
$p_7$ & 127.5225295696610033 & -90.4490412959934957 & -83.1783077030789002 & 34.0930433392580028 \\
\hline
\hline
\end{tabular}

\vspace{0.5cm}

  \begin{tabular}{ r  r }
    \hline
    \hline
    \textsc{parameter} & \textsc{value} \\
    \hline
%    ${m_H}$ \qquad & 125.0 \\
    ${m_W}$ \qquad & 80.419 GeV \\
    ${w_W}$ \qquad & 2.0476 GeV \\
%    ${m_Z}$ \qquad & 91.188 \\
%    ${m_t}$ \qquad & 171.2 \\
%    ${m_b}$ \qquad & 0.0 \\
    ${N_f}$ \qquad & 5 \\
%    ${N_{f,\textrm{gen}}}$ \qquad & 2 \\
    $\mu$ \qquad & 1000.0 GeV\\
    \hline
    \hline
\end{tabular}
\hspace{2.5cm}
  \begin{tabular}{ r  r r}
    \hline
    \hline
    & ${d}{\bar{u}}\rightarrow{\bar{\nu}_e}{e^-}{g}{g}{g}$  & Ref.~\cite{Badger:2012pg}\\
    \hline
$a_0$     \qquad &   \underline{-91.17720939046}11438 &  -91.17720939055536 \\
$a_{-1}$  \qquad &   \underline{-57.6891244440692}361  & -57.68912444409629\\
$a_{-2}$  \qquad &   \underline{-11.666666666666}8277  & -11.66666666666660 \\
\hline
    \hline
\end{tabular}
}%
\caption{Benchmark point for the subprocess $d(p_{1})\bar{u}(p_{2})\rightarrow \bar{\nu}_e(p_3)e^-(p_4)g(p_5)g(p_6)g(p_7)$.}
\label{tab:udenegggpsp}
\end{center}
%}
\end{table}

\end*

\subsection{$p\,p\rightarrow Z+3$ jets}
\paragraph*{Partonic process: ${d}{\bar{d}}\rightarrow{e^+}{e^-}{g}{g}{g}$}
~\\
The finite part for this process is given in CDR and was compared to the new version of \njet~\cite{Badger:2012pg}. We also find agreement in the phase space point given by the \blackhat{}
Collaboration~\cite{Berger:2010vm}.
\begin{table}[ht]
%\TABLE{
%\centering
\begin{center}
{\footnotesize
\begin{tabular}{c c c c c}
\hline 
\hline
\textsc{particle} & $E$& $p_x$ & $p_y$ & $p_z$ \\ 
\hline
$p_1$ & 500.0000000000000000 & 0.0000000000000000 & 0.0000000000000000 & 500.0000000000000000 \\
$p_2$ & 500.0000000000000000 & 0.0000000000000000 & 0.0000000000000000 & -500.0000000000000000 \\
$p_3$ & 414.1300683745429865 & 232.1455649459389861 & 332.7544367808189918 & -82.9857518524426041 \\
$p_4$ & 91.8751521026383955 & -43.3570226791010995 & -24.0058236140056991 & 77.3623460793434958 \\
$p_5$ & 86.3540681437814044 & -15.2133893202618005 & 37.6335512949163018 & -76.2187226821854011 \\
$p_6$ & 280.1181818093759830 & -83.1261116505822031 & -263.2038567586509998 & 47.7490851160265990 \\
$p_7$ & 127.5225295696610033 & -90.4490412959934957 & -83.1783077030789002 & 34.0930433392580028 \\
\hline
\hline
\end{tabular}

\vspace{0.5cm}

  \begin{tabular}{ r  r }
    \hline
    \hline
    \textsc{parameter} & \textsc{value} \\
    \hline
%    ${m_H}$ \qquad & 125.0 \\
    ${m_W}$ \qquad & 80.419 GeV\\
    ${m_Z}$ \qquad & 91.188 GeV\\
    ${w_Z}$ \qquad & 2.4414 GeV\\
%    ${m_t}$ \qquad & 171.2 \\
    %${m_b}$ \qquad & 0.0 \\
    ${N_f}$ \qquad & 5 \\
%    ${N_{f,\textrm{gen}}}$ \qquad & 1 \\
    $\mu$ \qquad & 1000.0 GeV \\
    \hline
    \hline
\end{tabular}
\hspace{2.0cm}
  \begin{tabular}{ r  r  r}
    \hline
    \hline
    & ${d}{\bar{d}}\rightarrow{e^+}{e^-}{g}{g}{g}$ & Ref.~\cite{Badger:2012pg} \\
    \hline
$a_0$     \qquad &   \underline{-91.04632912778}14761 & -91.04632919538757\\
$a_{-1}$  \qquad &   \underline{-57.687671748094}1892 & -57.68767174883881\\
$a_{-2}$  \qquad &   \underline{-11.666666666666}9485 & -11.66666666666667\\
\hline
    \hline
\end{tabular}
}%
\caption{Benchmark point for the subprocess $d(p_{1})\bar{d}(p_{2})\rightarrow e^+(p_3)e^-(p_4)g(p_5)g(p_6)g(p_7)$.}
\label{tab:ddeegggpsp}
\end{center}
%}
\end{table}

\newpage

\subsection{${p}\,{p}\rightarrow{Z}\,{Z}\,{Z}+1$ jet}
\paragraph*{Partonic process: ${u}{\bar{u}}\rightarrow{Z}{Z}{Z}{g}$}
~\\
\begin{table}[ht]
\begin{center}
{\footnotesize
\begin{tabular}{c c c c c}
\hline 
\hline
\textsc{particle} & $E$& $p_x$ & $p_y$ & $p_z$ \\ 
\hline
$p_1$ & 250.0000000000000000 & 0.0000000000000000 & 0.0000000000000000 & 250.0000000000000000 \\
$p_2$ & 250.0000000000000000 & 0.0000000000000000 & 0.0000000000000000 & -250.0000000000000000 \\
$p_3$ & 98.2984277476074197 & 25.7992145382080409 & -12.3240263933454042 & 23.0200218627010820 \\
$p_4$ & 178.4558180449861595 & -120.4664227955374400 & -73.7638561773535599 & -59.8166791207833768 \\
$p_5$ & 142.0905125919531145 & 50.7287365223860434 & 91.2424257480069656 & -31.2402050144644221 \\
$p_6$ & 81.1552416154533205 & 43.9384717349430645 & -5.1545431773078745 & 68.0368622725466565 \\
\hline
\hline
\end{tabular}

\vspace{0.5cm}

  \begin{tabular}{ r  r }
    \hline
    \hline
    \textsc{parameter} & \textsc{value} \\
    \hline
 %   ${m_H}$ \qquad & 125.0 \\
    ${m_W}$ \qquad & 80.376 GeV\\
    ${m_Z}$ \qquad & 91.1876 GeV\\
%    ${m_t}$ \qquad & 171.2 \\
  %  ${m_b}$ \qquad & 4.2 \\
    ${N_f}$ \qquad & 5 \\
 %   ${N_{f,\textrm{gen}}}$ \qquad & 5 \\
    $\mu$ \qquad & 500.0 GeV\\
    \hline
    \hline
\end{tabular}
\hspace{2.5cm}
  \begin{tabular}{ r  r }
    \hline
    \hline
    & ${u}{\bar{u}}\rightarrow{Z}{Z}{Z}{g}$ \\
    \hline
$a_0$     \qquad &   \underline{-18.227468766}9522883 \\
$a_{-1}$  \qquad &   \underline{-22.38323488318}61125 \\
$a_{-2}$  \qquad &   \underline{-5.666666666667}0755 \\
\hline
    \hline
\end{tabular}
}%
\caption{Benchmark point for the subprocess $u(p_{1})\bar{u}(p_{2})\rightarrow Z(p_3)Z(p_4)Z(p_5)g(p_6)$.}
%\label{tab:psp}
\end{center}
%}
\end{table}

%\newpage
\subsection{${p}\,{p}\rightarrow{W}\,{W}\,{Z}+1$ jet}
\paragraph*{Partonic process: ${u}{\bar{u}}\rightarrow{W^+}{W^-}{Z}{g}$}
~\\
\begin{table}[ht]
%\TABLE{
%\centering
\begin{center}
{\footnotesize
\begin{tabular}{c c c c c}
\hline 
\hline
\textsc{particle} & $E$& $p_x$ & $p_y$ & $p_z$ \\ 
\hline
$p_1$ & 250.0000000000000000 & 0.0000000000000000 & 0.0000000000000000 & 250.0000000000000000 \\
$p_2$ & 250.0000000000000000 & 0.0000000000000000 & 0.0000000000000000 & -250.0000000000000000 \\
$p_3$ & 89.1058761118447791 & 27.0348198946258513 & -12.9142626969235721 & 24.1225229592474193 \\
$p_4$ & 179.7207629659015140 & -126.2359378733789299 & -77.2966387614043384 & -62.6814876216509802 \\
$p_5$ & 146.1313400758695593 & 53.1582949292336409 & 95.6123118862003167 & -32.7363964816230890 \\
$p_6$ & 85.0420208463841476 & 46.0428230495191357 & -5.4014104278722739 & 71.2953611440265860 \\
\hline
\hline
\end{tabular}

\vspace{0.5cm}

  \begin{tabular}{ r  r }
    \hline
    \hline
    \textsc{parameter} & \textsc{value} \\
    \hline
 %   ${m_H}$ \qquad & 125.0 \\
    ${m_W}$ \qquad & 80.376 GeV\\
    ${m_Z}$ \qquad & 91.1876 GeV\\
%    ${m_t}$ \qquad & 171.2 \\
  %  ${m_b}$ \qquad & 4.2 \\
    ${N_f}$ \qquad & 5 \\
 %   ${N_{f,\textrm{gen}}}$ \qquad & 5 \\
    $\mu$ \qquad & 500.0 GeV\\
    \hline
    \hline
\end{tabular}
\hspace{2.5cm}
  \begin{tabular}{ r  r }
    \hline
    \hline
    & ${u}{\bar{u}}\rightarrow{W^+}{W^-}{Z}{g}$ \\
    \hline
$a_0$     \qquad &   \underline{-18.005030590}6438837 \\
$a_{-1}$  \qquad &   \underline{-22.10254520117}81987 \\
$a_{-2}$  \qquad &   \underline{-5.66666666666666}61 \\
\hline
    \hline
\end{tabular}
}%
\caption{Benchmark point for the subprocess $u(p_{1})\bar{u}(p_{2})\rightarrow W^+(p_3)W^-(p_4)Z(p_5)g(p_6)$.}
%\label{tab:psp}
\end{center}
%}
\end{table}

\newpage

\subsection{${p}\,{p}\rightarrow{W}\,{Z}\,{Z}+1$ jet}
\paragraph*{Partonic process: ${u}{\bar{d}}\rightarrow{W^+}{Z}{Z}{g}$}
~\\
\begin{table}[ht]
%\TABLE{
%\centering
\begin{center}
{\footnotesize
\begin{tabular}{c c c c c}
\hline 
\hline
\textsc{particle} & $E$& $p_x$ & $p_y$ & $p_z$ \\ 
\hline
$p_1$ & 250.0000000000000000 & 0.0000000000000000 & 0.0000000000000000 & 250.0000000000000000 \\
$p_2$ & 250.0000000000000000 & 0.0000000000000000 & 0.0000000000000000 & -250.0000000000000000 \\
$p_3$ & 88.8514948513947331 & 26.6180234830689777 & -12.7151632255396141 & 23.7506254934495686 \\
$p_4$ & 182.6577199093957518 & -124.2897556491134168 & -76.1049547854528043 & -61.7151257515301381 \\
$p_5$ & 144.7598590943426586 & 52.3387523298249846 & 94.1382547757718982 & -32.2316987387113372 \\
$p_6$ & 83.7309261448668849 & 45.3329798362191525 & -5.3181367647793465 & 70.1961989967918498 \\
\hline
\hline
\end{tabular}

\vspace{0.5cm}

  \begin{tabular}{ r  r }
    \hline
    \hline
    \textsc{parameter} & \textsc{value} \\
    \hline
%    ${m_H}$ \qquad & 125.0 \\
    ${m_W}$ \qquad & 80.376 GeV\\
    ${m_Z}$ \qquad & 91.1876 GeV\\
  %  ${m_t}$ \qquad & 171.2 \\
   % ${m_b}$ \qquad & 0.0 \\
    ${N_f}$ \qquad & 5 \\
   % ${N_{f,\textrm{gen}}}$ \qquad & 2 \\
   % $\mu^2$ \qquad & 250000.0 \\
    $\mu$ \qquad & 500.0 GeV \\
    \hline 
    \hline
\end{tabular}
\hspace{2.5cm}
  \begin{tabular}{ r  r }
    \hline
    \hline
    & ${u}{\bar{d}}\rightarrow{W^+}{Z}{Z}{g}$ \\
    \hline
$a_0$     \qquad &   \underline{-16.67196386149}81740 \\
$a_{-1}$  \qquad &   \underline{-22.19576780110}10735 \\
$a_{-2}$  \qquad &   \underline{-5.666666666667}0213 \\
\hline
    \hline
\end{tabular}
}%
\caption{Benchmark point for the subprocess $u(p_{1})\bar{d}(p_{2})\rightarrow W^+(p_3)Z(p_4)Z(p_5)g(p_6)$.}
\label{tab:psp}
\end{center}
%}
\end{table}

\subsection{${p}\,{p}\rightarrow{W}\,{W}\,{W}+1$ jet}
\paragraph*{Partonic process: ${u}{\bar{d}}\rightarrow{W^+}{W^-}{W^+}{g}$}
~\\
\begin{table}[ht]
%\TABLE{
%\centering
\begin{center}
{\footnotesize
\begin{tabular}{c c c c c}
\hline 
\hline
\textsc{particle} & $E$& $p_x$ & $p_y$ & $p_z$ \\ 
\hline
$p_1$ & 250.0000000000000000 & 0.0000000000000000 & 0.0000000000000000 & 250.0000000000000000 \\
$p_2$ & 250.0000000000000000 & 0.0000000000000000 & 0.0000000000000000 & -250.0000000000000000 \\
$p_3$ & 89.4258007278425993 & 27.5517039326065500 & -13.1611730250244197 & 24.5837262194154818 \\
$p_4$ & 182.4747913234621421 & -128.6494675939613614 & -78.7744883983191500 & -63.8799073098456347 \\
$p_5$ & 141.4314519821789986 & 54.1746388235997784 & 97.4403424793774207 & -33.3622900835894285 \\
$p_6$ & 86.6679559665162316 & 46.9231248377547274 & -5.5046810560337240 & 72.6584711740195104 \\
\hline
\hline
\end{tabular}

\vspace{0.5cm}

  \begin{tabular}{ r  r }
    \hline
    \hline
    \textsc{parameter} & \textsc{value} \\
    \hline
%    ${m_H}$ \qquad & 125.0 \\
    ${m_W}$ \qquad & 80.376 GeV\\
    ${m_Z}$ \qquad & 91.1876 GeV\\
  %  ${m_t}$ \qquad & 171.2 \\
   % ${m_b}$ \qquad & 0.0 \\
    ${N_f}$ \qquad & 5 \\
   % ${N_{f,\textrm{gen}}}$ \qquad & 2 \\
   % $\mu^2$ \qquad & 250000.0 \\
    $\mu$ \qquad & 500.0 GeV \\
    \hline
    \hline
\end{tabular}
\hspace{2.5cm}
  \begin{tabular}{ r  r }
    \hline
    \hline
    & ${u}{\bar{d}}\rightarrow{W^+}{W^-}{W^+}{g}$ \\
    \hline
$a_0$     \qquad &   \underline{-15.8859769338}002099 \\
$a_{-1}$  \qquad &   \underline{-21.9889128719}035618 \\
$a_{-2}$  \qquad &   \underline{-5.6666666666}864467 \\
\hline
    \hline
\end{tabular}
}%
\caption{Benchmark point for the subprocess $u(p_{1})\bar{d}(p_{2})\rightarrow W^+(p_3)W^-(p_4)W^+(p_5)g(p_6)$.}
\label{tab:psp}
\end{center}
%}
\end{table}

\newpage

\subsection{${p}\,{p}\rightarrow{Z}\,{Z}\,{Z}\,{Z}$}
\paragraph*{Partonic process: ${u}\,{\bar{u}}\rightarrow{Z}\,{Z}\,{Z}\,{Z}$}
~\\
\begin{table}[ht]
%\TABLE{
%\centering
\begin{center}
{\footnotesize
\begin{tabular}{c c c c c}
\hline 
\hline
\textsc{particle} & $E$& $p_x$ & $p_y$ & $p_z$ \\ 
\hline
$p_1$ & 250.0000000000000000 & 0.0000000000000000 & 0.0000000000000000 & 250.0000000000000000 \\
$p_2$ & 250.0000000000000000 & 0.0000000000000000 & 0.0000000000000000 & -250.0000000000000000 \\
$p_3$ & 96.3863867610220666 & 21.9480082963376795 & -10.4843437432888980 & 19.5836826767025762 \\
$p_4$ & 159.2027435644542095 & -102.4836644941604078 & -62.7526750844588079 & -50.8874782857443790 \\
$p_5$ & 130.0351856078737001 & 43.1561483551038094 & 77.6221118797800074 & -26.5767889104262487 \\
$p_6$ & 114.3756840666499812 & 37.3795078427186667 & -4.3850930520321931 & 57.8805845194680018 \\
\hline
\hline
\end{tabular}

\vspace{0.5cm}

  \begin{tabular}{ r  r }
    \hline
    \hline
    \textsc{parameter} & \textsc{value} \\
    \hline
%    ${m_H}$ \qquad & 125.0 \\
    ${m_W}$ \qquad & 80.376 GeV\\
    ${m_Z}$ \qquad & 91.1876 GeV\\
  %  ${m_t}$ \qquad & 171.2 \\
   % ${m_b}$ \qquad & 0.0 \\
    ${N_f}$ \qquad & 5 \\
   % ${N_{f,\textrm{gen}}}$ \qquad & 2 \\
   % $\mu^2$ \qquad & 250000.0 \\
    $\mu$ \qquad & 500.0 GeV \\
    \hline
    \hline
\end{tabular}
\hspace{2.5cm}
  \begin{tabular}{ r  r }
    \hline
    \hline
    & ${u}{\bar{u}}\rightarrow{Z}{Z}{Z}{Z}$ \\
    \hline
$a_0$     \qquad &   \underline{10.00102685}60339206 \\
$a_{-1}$  \qquad &   \underline{-3.9999999999}613696 \\
$a_{-2}$  \qquad &   \underline{-2.6666666666665}884 \\
\hline
    \hline
\end{tabular}
}%
\caption{Benchmark point for the subprocess $u(p_{1})\bar{u}(p_{2})\rightarrow Z(p_3)Z(p_4)Z(p_5)Z(p_6)$.}
\label{tab:psp}
\end{center}
%}
\end{table}

\subsection{${p}\,{p}\rightarrow{W}\,{W}\,{W}\,{W}$}
\paragraph*{Partonic process: ${u}{\bar{u}}\rightarrow{W^+}{W^-}{W^+}{W^-}$}
~\\
\begin{table}[ht]
%\TABLE{
%\centering
\begin{center}
{\footnotesize
\begin{tabular}{c c c c c}
\hline 
\hline
\textsc{particle} & $E$& $p_x$ & $p_y$ & $p_z$ \\ 
\hline
$p_1$ & 250.0000000000000000 & 0.0000000000000000 & 0.0000000000000000 & 250.0000000000000000 \\
$p_2$ & 250.0000000000000000 & 0.0000000000000000 & 0.0000000000000000 & -250.0000000000000000 \\
$p_3$ & 87.7920438094441096 & 24.8207950462383629 & -11.8566452013738353 & 22.1470015595954024 \\
$p_4$ & 168.0494737136866092 & -115.8978071108833205 & -70.9664068759461202 & -57.5481680112689915 \\
$p_5$ & 132.1031656516532848 & 48.8048800986569518 & 87.7821123452778238 & -30.0554392738635485 \\
$p_6$ & 112.0553168252159821 & 42.2721319659877182 & -4.9590602679577351 & 65.4566057255370879 \\
\hline
\hline
\end{tabular}

\vspace{0.5cm}

  \begin{tabular}{ r  r }
    \hline
    \hline
    \textsc{parameter} & \textsc{value} \\
    \hline
  %  ${m_H}$ \qquad & 125.0 \\
     ${m_W}$ \qquad & 80.376 GeV\\
    ${m_Z}$ \qquad & 91.1876 GeV\\
  %  ${m_t}$ \qquad & 171.2 \\
   % ${m_b}$ \qquad & 0.0 \\
    ${N_f}$ \qquad & 5 \\
   % ${N_{f,\textrm{gen}}}$ \qquad & 2 \\
   % $\mu^2$ \qquad & 250000.0 \\
    $\mu$ \qquad & 500.0 GeV \\    \hline
    \hline
\end{tabular}
\hspace{2.5cm}
  \begin{tabular}{ r  r }
    \hline
    \hline
    & ${u}{\bar{u}}\rightarrow{W^+}{W^-}{W^+}{W^-}$ \\
    \hline
$a_0$     \qquad &   \underline{7.85563273962}45011 \\
$a_{-1}$  \qquad &   \underline{-3.99999999999}81126 \\
$a_{-2}$  \qquad &   \underline{-2.666666666666}9747 \\
\hline
    \hline
\end{tabular}
}%
\caption{Benchmark point for the subprocess $u(p_{1})\bar{u}(p_{2})\rightarrow W^+(p_3)W^-(p_4)W^+(p_5)W^-(p_6)$.}
\label{tab:psp}
\end{center}
%}
\end{table}

\newpage
\subsection{$p\,p\rightarrow t\,\bar{t}\,b\,\bar{b}$}
\paragraph*{Partonic process: ${d}{\bar{d}}\rightarrow{t}{\bar{t}}{b}{\bar{b}}$}
~\\
\begin{table}[ht]
%\TABLE{
%\centering
\begin{center}
{\footnotesize
\begin{tabular}{c c c c c}
\hline 
\hline
\textsc{particle} & $E$& $p_x$ & $p_y$ & $p_z$ \\ 
\hline
$p_1$ & 250.0000000000000000 & 0.0000000000000000 & 0.0000000000000000 & 250.0000000000000000 \\
$p_2$ & 250.0000000000000000 & 0.0000000000000000 & 0.0000000000000000 & -250.0000000000000000 \\
$p_3$ & 213.1527786548405459 & -26.8458616570582116 & -117.8628562648380296 & -38.8884799556846303 \\
$p_4$ & 220.5509350311838546 & 61.9684848664477599 & 92.5635212096267281 & 83.2176445698946168 \\
$p_5$ & 42.2714703981682263 & -16.8075489037092431 & 24.9923105627744704 & -29.3620149548096769 \\
$p_6$ & 24.0248159158073982 & -18.3150743056803300 & 0.3070244924368429 & -14.9671496594002438 \\
\hline
\hline
\end{tabular}

\vspace{0.5cm}

  \begin{tabular}{ r  r }
    \hline
    \hline
    \textsc{parameter} & \textsc{value} \\
    \hline
%    ${m_H}$ \qquad & 125.0 \\
%    ${m_W}$ \qquad & 80.376 GeV\\
%    ${m_Z}$ \qquad & 91.1876 GeV\\
    ${m_t}$ \qquad & 171.2 GeV\\
    ${m_b}$ \qquad & 4.2 GeV\\
    ${N_f}$ \qquad & 4 \\
%    ${N_{f,\textrm{gen}}}$ \qquad & 4 \\
    $\mu$ \qquad & 500.0 GeV\\
    \hline
    \hline
\end{tabular}
\hspace{2.5cm}
  \begin{tabular}{ r  r }
    \hline
    \hline
    & ${d}{\bar{d}}\rightarrow{t}{\bar{t}}{b}{\bar{b}}$ \\
    \hline
$a_0$     \qquad &   \underline{154.64756730066}05973 \\
$a_{-1}$  \qquad &   \underline{2.740905091457}7211 \\
$a_{-2}$  \qquad &   \underline{-2.66666666666666}83 \\
\hline
    \hline
\end{tabular}
}%
\caption{Benchmark point for the subprocess $d(p_{1})\bar{d}(p_{2})\rightarrow t(p_3)\bar{t}(p_4)b(p_5)\bar{b}(p_6)$.}
\label{tab:ddttbbpsp}
\end{center}
%}
\end{table}

%\newpage
\paragraph*{Partonic process: ${g}{g}\rightarrow{t}{\bar{t}}{b}{\bar{b}}$}
~\\
\begin{table}[ht]
%\TABLE{
%\centering
\begin{center}
{\footnotesize
\begin{tabular}{c c c c c}
\hline 
\hline
\textsc{particle} & $E$& $p_x$ & $p_y$ & $p_z$ \\ 
\hline
$p_1$ & 250.0000000000000000 & 0.0000000000000000 & 0.0000000000000000 & 250.0000000000000000 \\
$p_2$ & 250.0000000000000000 & 0.0000000000000000 & 0.0000000000000000 & -250.0000000000000000 \\
$p_3$ & 194.0670578462199387 & -60.6594324624948058 & -47.3641590248774236 & -49.2915067154802884 \\
$p_4$ & 172.4499944124030151 & -15.6689752760792977 & -7.1446619651393677 & -11.5324581958636152 \\
$p_5$ & 61.9093840678718479 & 12.0715208460656545 & 23.6785835371366993 & 55.7560301833003820 \\
$p_6$ & 71.5735636735052054 & 64.2568868925084331 & 30.8302374528801408 & 5.0679347280435243 \\
\hline
\hline
\end{tabular}

\vspace{0.5cm}

  \begin{tabular}{ r  r }
    \hline
    \hline
    \textsc{parameter} & \textsc{value} \\
    \hline
%    ${m_H}$ \qquad & 125.0 \\
%    ${m_W}$ \qquad & 80.376 GeV\\
%    ${m_Z}$ \qquad & 91.1876 GeV\\
    ${m_t}$ \qquad & 171.2 GeV\\
    ${m_b}$ \qquad & 4.2 GeV\\
    ${N_f}$ \qquad & 4 \\
%    ${N_{f,\textrm{gen}}}$ \qquad & 4 \\
    $\mu$ \qquad & 500.0 GeV\\
    \hline
    \hline
\end{tabular}
\hspace{2.5cm}
  \begin{tabular}{ r  r }
    \hline
    \hline
    & ${g}{g}\rightarrow{t}{\bar{t}}{b}{\bar{b}}$ \\
    \hline
$a_0$     \qquad &   \underline{165.12500385}52732349 \\
$a_{-1}$  \qquad &   \underline{-3.447272593}0136989 \\
$a_{-2}$  \qquad &   \underline{-6.000000000000}1563 \\
\hline
    \hline
\end{tabular}
}%
\caption{Benchmark point for the subprocess $g(p_{1})g(p_{2})\rightarrow t(p_3)\bar{t}(p_4)b(p_5)\bar{b}(p_6)$.}
\label{tab:ggttbbpsp}
\end{center}
%}
\end{table}

\newpage
\subsection{$p\,p\rightarrow t\,\bar{t}+2$ jets}
\paragraph*{Partonic process: ${g}{g}\rightarrow {t}{\bar{t}}{g}{g}$}
~\\
The finite part for this process is given in CDR and was compared with Ref.~\cite{Bevilacqua:2011aa}.
\begin{table}[ht]
%\TABLE{
%\centering
\begin{center}
{\footnotesize
\begin{tabular}{c c c c c}
\hline 
\hline
\textsc{particle} & $E$& $p_x$ & $p_y$ & $p_z$ \\ 
\hline
$p_1$ & 2424.7465026975200999 & 0.0000000000000000 & 0.0000000000000000 & 2424.7465026975200999 \\
$p_2$ & 2424.7465026975200999 & 0.0000000000000000 & 0.0000000000000000 & -2424.7465026975200999 \\
$p_3$ & 881.9042263139403985 & -715.3340594013137661 & -475.1625187999429158 & 101.1925816377931966 \\
$p_4$ & 343.4841188963524132 & -24.1478321960174789 & -6.3283366075706340 & 295.5085181344487069 \\
$p_5$ & 1673.4634600426329598 & 21.8782679485017297 & 1000.4115637957629588 & 1341.3344089052341133 \\
$p_6$ & 1950.6412001421140303 & 717.6036236488295117 & -518.9207083882492952 & -1738.0355086774759457 \\
\hline
\hline
\end{tabular}

\vspace{0.5cm}

  \begin{tabular}{ r  r }
    \hline
    \hline
    \textsc{parameter} & \textsc{value} \\
    \hline
%    ${m_H}$ \qquad & 125.0 \\
%    ${m_W}$ \qquad & 80.376 \\
%    ${m_Z}$ \qquad & 91.1876 \\
    ${m_t}$ \qquad & 173.3 GeV \\
%    ${m_b}$ \qquad & 0.0 \\
    ${N_f}$ \qquad & 5 \\
%    ${N_{f,\textrm{gen}}}$ \qquad & 1 \\
    $\mu$ \qquad &  173.3 GeV\\
    \hline
    \hline
\end{tabular}
\hspace{2.5cm}
  \begin{tabular}{ r  r r}
    \hline
    \hline
    & ${g}{g}\rightarrow{t}{\bar{t}}{g}{g}$ & Ref.~\cite{Bevilacqua:2011aa}\\
    \hline
$a_0$     \qquad &   \underline{-93.98254280686}26394 & -93.98254280655584 \\
$a_{-1}$  \qquad &   \underline{47.0991735298}819236 & 47.0991735300671 \\
$a_{-2}$  \qquad &   \underline{-11.99999999999}47171 & -11.999999999999874\\
\hline
    \hline
\end{tabular}
}%
\caption{Benchmark point for the subprocess $g(p_{1})g(p_{2})\rightarrow t(p_3)\bar{t}(p_4)g(p_5)g(p_6)$.}
\label{tab:ggttbarggpsp}
\end{center}
%}
\end{table}

\subsection{$p\,p\rightarrow Z\,b\,\bar{b}+1$ jet}
\paragraph*{Partonic process: ${u}{g}\rightarrow{u}{e^+}{e^-}{b}{\bar{b}}$}
~\\ 
\begin{table}[ht]
%\TABLE{
%\centering
\begin{center}
{\footnotesize
\begin{tabular}{c c c c c}
\hline 
\hline
\textsc{particle} & $E$& $p_x$ & $p_y$ & $p_z$ \\ 
\hline
$p_1$ & 250.0000000000000000 & 0.0000000000000000 & 0.0000000000000000 & 250.0000000000000000 \\
$p_2$ & 250.0000000000000000 & 0.0000000000000000 & 0.0000000000000000 & -250.0000000000000000 \\
$p_3$ & 34.5908278264605187 & 21.2091909896352142 & -9.4555401980608202 & 25.6376353599131122 \\
$p_4$ & 166.7223525775519022 & -156.7623134542972991 & -48.7827423195217946 & -29.0200617028515602 \\
$p_5$ & 111.6942046513332798 & 30.9750523871488710 & 106.1302756897373314 & -15.8904394000814051 \\
$p_6$ & 84.2714416207739418 & 35.0918815486497024 & 1.4231216042880672 & 76.4890217424595988 \\
$p_7$ & 102.7211733238803646 & 69.4861885288632521 & -49.3151147764427051 & -57.2161559994397777 \\
\hline
\hline
\end{tabular}

\vspace{0.5cm}

  \begin{tabular}{ r  r }
    \hline
    \hline
    \textsc{parameter} & \textsc{value} \\
    \hline
    ${m_W}$ \qquad & 80.376 GeV\\
    ${w_W}$ \qquad & 2.124 GeV \\
%    ${m_Z}$ \qquad & 91.1876 GeV\\
%    ${m_t}$ \qquad & 171.2 \\
    ${m_b}$ \qquad & 4.2 GeV\\
    ${N_f}$ \qquad & 4 \\
%    ${N_{f,\textrm{gen}}}$ \qquad & 2 \\
    $\mu$ \qquad & 500.0 GeV\\
    \hline
    \hline
\end{tabular}
\hspace{2.5cm}
  \begin{tabular}{ r  r }
    \hline
    \hline
    & ${u}{g}\rightarrow{u}{e^+}{e^-}{b}{\bar{b}}$ \\
    \hline
$a_0$     \qquad &   \underline{145.370895468}0396630 \\
$a_{-1}$  \qquad &   \underline{-8.36795126937}08471 \\
$a_{-2}$  \qquad &   \underline{-5.666666666667}5392 \\
\hline
    \hline
\end{tabular}
}%
\caption{Benchmark point for the subprocess $u(p_{1})g(p_{2})\rightarrow u(p_3)e^+(p_4)e^-(p_5)b(p_6)\bar{b}(p_7)$.}
\label{tab:psp}
\end{center}
%}
\end{table}

\newpage

\subsection{$p\,p\rightarrow W\,b\,\bar{b}+1$ jet}
\paragraph*{Partonic process: ${u}{\bar{d}}\rightarrow{e^+}{\nu_e}{b}{\bar{b}}{g}$}
~\\
\begin{table}[ht]
%\TABLE{
%\centering
\begin{center}
{\footnotesize
\begin{tabular}{c c c c c}
\hline 
\hline
\textsc{particle} & $E$& $p_x$ & $p_y$ & $p_z$ \\ 
\hline
$p_1$ & 250.0000000000000000 & 0.0000000000000000 & 0.0000000000000000 & 250.0000000000000000 \\
$p_2$ & 250.0000000000000000 & 0.0000000000000000 & 0.0000000000000000 & -250.0000000000000000 \\
$p_3$ & 93.4300963683620580 & -16.6492363753179085 & -37.4579803162897420 & -83.9576413803095818 \\
$p_4$ & 119.9994454272237192 & 4.7053605726301706 & -100.6826015733804809 & 65.1209660949429150 \\
$p_5$ & 57.9859979994296282 & 9.1381348721238638 & -4.6735873988293006 & 56.9156220722767259 \\
$p_6$ & 104.9559149323387999 & 87.3260122226470514 & 54.3049824548373437 & -20.5728109201014071 \\
$p_7$ & 123.6285452726457805 & -84.5202712920831658 & 88.5091868336622554 & -17.5061358668087337 \\
\hline
\hline
\end{tabular}

\vspace{0.5cm}

  \begin{tabular}{ r  r }
    \hline
    \hline
    \textsc{parameter} & \textsc{value} \\
    \hline
%    ${m_H}$ \qquad & 125.0 \\
    ${m_W}$ \qquad & 80.376 GeV\\
    ${w_W}$ \qquad & 2.124 GeV \\
%    ${m_Z}$ \qquad & 91.1876 GeV\\
%    ${m_t}$ \qquad & 171.2 \\
    ${m_b}$ \qquad & 4.2 GeV\\
    ${N_f}$ \qquad & 4 \\
%    ${N_{f,\textrm{gen}}}$ \qquad & 2 \\
    $\mu$ \qquad & 500.0 GeV\\
    \hline
    \hline
\end{tabular}
\hspace{2.5cm}
  \begin{tabular}{ r  r }
    \hline
    \hline
    & ${u}{\bar{d}}\rightarrow{e^+}{\nu_e}{b}{\bar{b}}{g}$ \\
    \hline
$a_0$     \qquad &   \underline{129.953855486477}1243 \\
$a_{-1}$  \qquad &   \underline{-5.338568370118}9715 \\
$a_{-2}$  \qquad &   \underline{-5.666666666666}8695 \\
\hline
    \hline
\end{tabular}
}%
\caption{Benchmark point for the subprocess $u(p_{1})\bar{d}(p_{2})\rightarrow e^+(p_3)\nu_e(p_4)b(p_5)\bar{b}(p_6)g(p_7)$.}
\label{tab:udenebbgpsp}
\end{center}
%}
\end{table}

\subsection{$p\,p\rightarrow W\,b\,\bar{b}+2$ jets}
\paragraph*{Partonic process: ${u}{\bar{d}}\rightarrow{e^+}{\nu_e}{b}{\bar{b}}{d}{\bar{d}}$}
~\\
%\glnote{Shall we specify that there is not top in the loops??}
\begin{table}[ht]
%\TABLE{
%\centering
\begin{center}
{\footnotesize
\begin{tabular}{c c c c c}
\hline 
\hline
\textsc{particle} & $E$& $p_x$ & $p_y$ & $p_z$ \\ 
\hline
$p_1$ & 250.0000000000000000 & 0.0000000000000000 & 0.0000000000000000 & 250.0000000000000000 \\
$p_2$ & 250.0000000000000000 & 0.0000000000000000 & 0.0000000000000000 & -250.0000000000000000 \\
$p_3$ & 125.6965187314100234 & -17.0943040170214537 & -113.3597903541534748 & 51.5456838370753374 \\
$p_4$ & 72.1993434263444129 & 25.0205029393394440 & 42.4573677065765622 & 52.7644913865970722 \\
$p_5$ & 130.5494441454287085 & -59.2470822889473183 & 116.2307891878907924 & -2.3883575291830641 \\
$p_6$ & 52.9433261580396106 & 45.1283306603629413 & -14.5296908876010313 & -23.1878769876905828 \\
$p_7$ & 99.1517346871049057 & -5.9484899951818377 & -31.1690273964595583 & -93.9370730297576273 \\
$p_8$ & 19.4596328516722785 & 12.1410427014482281 & 0.3703517437466808 & 15.2031323229588882 \\
\hline
\hline
\end{tabular}

\vspace{0.5cm}

  \begin{tabular}{ r  r }
    \hline
    \hline
    \textsc{parameter} & \textsc{value} \\
    \hline
%    ${m_H}$ \qquad & 125.0 \\
    ${m_W}$ \qquad & 80.376 GeV\\
    ${w_W}$ \qquad & 2.124 GeV \\
%    ${m_Z}$ \qquad & 91.1876 GeV\\
%    ${m_t}$ \qquad & 171.2 \\
    ${m_b}$ \qquad & 4.2 GeV\\
    ${N_f}$ \qquad & 4 \\
%    ${N_{f,\textrm{gen}}}$ \qquad & 2 \\
    $\mu$ \qquad & 500.0 GeV\\
    \hline
    \hline
\end{tabular}
\hspace{2.5cm}
  \begin{tabular}{ r  r }
    \hline
    \hline
    & ${u}{\bar{d}}\rightarrow{e^+}{\nu_e}{b}{\bar{b}}{d}{\bar{d}}$ \\
    \hline
$a_0$     \qquad &   \underline{148.249956413826}0837 \\
$a_{-1}$  \qquad &   \underline{-7.727299512216}3879 \\
$a_{-2}$  \qquad &   \underline{-5.333333333333}1892 \\
\hline
    \hline
\end{tabular}
}%
\caption{Benchmark point for the subprocess $u(p_{1})\bar{d}(p_{2})\rightarrow e^+(p_3)\nu_e(p_4)b(p_5)\bar{b}(p_6)d(p_7)\bar{d}(p_8)$.}
\label{tab:udenebbddpsp}
\end{center}
%}
\end{table}

\newpage

\paragraph*{Partonic process: ${u}{\bar{d}}\rightarrow{e^+}{\nu_e}{b}{\bar{b}}{s}{\bar{s}}$}
~\\
\begin{table}[ht]
%\TABLE{
%\centering
\begin{center}
{\footnotesize
\begin{tabular}{c c c c c}
\hline 
\hline
\textsc{particle} & $E$& $p_x$ & $p_y$ & $p_z$ \\ 
\hline
$p_1$ & 250.0000000000000000 & 0.0000000000000000 & 0.0000000000000000 & 250.0000000000000000 \\
$p_2$ & 250.0000000000000000 & 0.0000000000000000 & 0.0000000000000000 & -250.0000000000000000 \\
$p_3$ & 1.8312125180161738 & 1.2481975210878733 & -0.3016228634824342 & -1.3055136470806736 \\
$p_4$ & 132.0663362603577298 & -16.6607423420677527 & 94.8432229336175965 & -90.3808602603960196 \\
$p_5$ & 121.6089450080674226 & -63.6699664483805989 & -102.6429343577717077 & 13.4780898076629523 \\
$p_6$ & 51.8272368161295987 & -33.3143054988286877 & -36.6615737768635270 & -14.6461098360448521 \\
$p_7$ & 124.2305315458477253 & 116.2047315474863467 & -2.8413585495376918 & 43.8361952698163435 \\
$p_8$ & 68.4357378515813224 & -3.8079147792972257 & 47.6042666140377335 & 49.0181986660422169 \\
\hline
\hline
\end{tabular}

\vspace{0.5cm}

  \begin{tabular}{ r  r }
    \hline
    \hline
    \textsc{parameter} & \textsc{value} \\
    \hline
%    ${m_H}$ \qquad & 125.0 \\
    ${m_W}$ \qquad & 80.376 GeV\\
    ${w_W}$ \qquad & 2.124 GeV \\    
%    ${m_Z}$ \qquad & 91.1876 GeV\\
%    ${m_t}$ \qquad & 171.2 \\
    ${m_b}$ \qquad & 4.2 GeV\\
    ${N_f}$ \qquad & 4 \\
%    ${N_{f,\textrm{gen}}}$ \qquad & 3 \\
    $\mu$ \qquad & 500.0 GeV\\
    \hline
    \hline
\end{tabular}
\hspace{2.5cm}
  \begin{tabular}{ r  r }
    \hline
    \hline
    & ${u}{\bar{d}}\rightarrow{e^+}{\nu_e}{b}{\bar{b}}{s}{\bar{s}}$ \\
    \hline
$a_0$     \qquad &   \underline{161.16563616777}29592 \\
$a_{-1}$  \qquad &   \underline{-1.327626226075}3209 \\
$a_{-2}$  \qquad &   \underline{-5.3333333333332}735 \\
\hline
    \hline
\end{tabular}
}%
\caption{Benchmark point for the subprocess $u(p_{1})\bar{d}(p_{2})\rightarrow e^+(p_3)\nu_e(p_4)b(p_5)\bar{b}(p_6)s(p_7)\bar{s}(p_8)$.}
\label{tab:udenebbddpsp}
\end{center}
%}
\end{table}

%\newpage
\paragraph*{Partonic process: ${u}{\bar{d}}\rightarrow{e^+}{\nu_e}{b}{\bar{b}}{g}{g}$}
~\\
\begin{table}[ht]
%\TABLE{
%\centering
\begin{center}
{\footnotesize
\begin{tabular}{c c c c c}
\hline 
\hline
\textsc{particle} & $E$& $p_x$ & $p_y$ & $p_z$ \\ 
\hline
$p_1$ & 250.0000000000000000 & 0.0000000000000000 & 0.0000000000000000 & 250.0000000000000000 \\
$p_2$ & 250.0000000000000000 & 0.0000000000000000 & 0.0000000000000000 & -250.0000000000000000 \\
$p_3$ & 118.4116290336479551 & 49.7302976872290259 & -105.1432905838027665 & -22.2058512007951201 \\
$p_4$ & 9.9876328289379046 & 7.4336546931814604 & 4.5250271451401858 & 4.9007873616352553 \\
$p_5$ & 113.2724870353399353 & -8.8682847027615033 & 110.4231681698792471 & 23.2614225044082019 \\
$p_6$ & 50.3648230617062964 & -38.8899369624844553 & -31.0689344437605435 & -6.4241355543198972 \\
$p_7$ & 105.2876197360011901 & -28.2113391151385819 & 10.7916624062129962 & 100.8620009592996638 \\
$p_8$ & 102.6758083043666687 & 18.8056083999740657 & 10.4723673063308507 & -100.3942240702281055 \\
\hline
\hline
\end{tabular}

\vspace{0.5cm}

  \begin{tabular}{ r  r }
    \hline
    \hline
    \textsc{parameter} & \textsc{value} \\
    \hline
%    ${m_H}$ \qquad & 125.0 \\
    ${m_W}$ \qquad & 80.376 GeV\\
    ${w_W}$ \qquad & 2.124 GeV \\    
%    ${m_Z}$ \qquad & 91.1876 GeV\\
%    ${m_t}$ \qquad & 171.2 \\
    ${m_b}$ \qquad & 4.2 GeV\\
    ${N_f}$ \qquad & 4 \\
%    ${N_{f,\textrm{gen}}}$ \qquad & 2 \\
    $\mu$ \qquad & 500.0 GeV\\
    \hline
    \hline
\end{tabular}
\hspace{2.5cm}
  \begin{tabular}{ r  r }
    \hline
    \hline
    & ${u}{\bar{d}}\rightarrow{e^+}{\nu_e}{b}{\bar{b}}{g}{g}$ \\
    \hline
$a_0$     \qquad &   \underline{64.89357705697}83301 \\
$a_{-1}$  \qquad &   \underline{-35.9610562256}753425 \\
$a_{-2}$  \qquad &   \underline{-8.666666666667}0285 \\
\hline
    \hline
\end{tabular}
}%
\caption{Benchmark point for the subprocess $u(p_{1})\bar{d}(p_{2})\rightarrow e^+(p_3)\nu_e(p_4)b(p_5)\bar{b}(p_6)g(p_7)g(p_8)$.}
\label{tab:udenebbggpsp}
\end{center}
%}
\end{table}

\newpage

\subsection{${p}\,{p}\rightarrow{W}\,{W}\,{b}\,{\bar b}$}
\paragraph*{Partonic process: ${d}{\bar{d}}\rightarrow{\nu_e}{e^+}{\bar{\nu}_\mu}{\mu^-}{b}{\bar{b}}$}
~\\
\begin{table}[ht]
%\TABLE{
%\centering
\begin{center}
{\footnotesize
\begin{tabular}{c c c c c}
\hline 
\hline
\textsc{particle} & $E$& $p_x$ & $p_y$ & $p_z$ \\ 
\hline
$p_1$ & 250.0000000000000000 & 0.0000000000000000 & 0.0000000000000000 & 250.0000000000000000 \\
$p_2$ & 250.0000000000000000 & 0.0000000000000000 & 0.0000000000000000 & -250.0000000000000000 \\
$p_3$ & 25.6649135887983881 & 15.7119189648809563 & -7.6256852614994211 & 18.8061776088085644 \\
$p_4$ & 154.5064489409292605 & -136.9768135536888565 & -46.2042967903236104 & -54.5413446032057010 \\
$p_5$ & 96.1603526761926730 & 22.3143342117606984 & 88.8440473781699325 & -29.2509685474019605 \\
$p_6$ & 61.0731578730670606 & 23.2528867537403734 & 2.3717946919774544 & 56.4234743714476039 \\
$p_7$ & 93.6199892430268648 & 58.9397665353324101 & -43.8226058551382351 & -57.9028973600646211 \\
$p_8$ & 68.9751376779857708 & 16.7579070879743455 & 6.4367458368138966 & 66.4655585304161036 \\
\hline
\hline
\end{tabular}

\vspace{0.5cm}

  \begin{tabular}{ r  r }
    \hline
    \hline
    \textsc{parameter} & \textsc{value} \\
    \hline
%    ${m_H}$ \qquad & 125.0 \\
    ${m_W}$ \qquad & 80.376 GeV\\
    ${m_Z}$ \qquad & 91.1876 GeV \\
    ${m_t}$ \qquad & 171.2 GeV\\
    ${m_b}$ \qquad & 4.2 GeV\\
    ${N_f}$ \qquad & 4 \\
%    ${N_{f,\textrm{gen}}}$ \qquad & 2 \\
    $\mu$ \qquad & 500.0 GeV\\
    \hline
    \hline
\end{tabular}
\hspace{2.5cm}
  \begin{tabular}{ r  r }
    \hline
    \hline
    & ${d}{\bar{d}}\rightarrow{\nu_e}{e^+}{\bar{\nu}_\mu}{\mu^-}{b}{\bar{b}}$ \\
    \hline
$a_0$     \qquad &   \underline{118.3990}066409585751 \\
$a_{-1}$  \qquad &   \underline{8.55743849262}30991 \\
$a_{-2}$  \qquad &   \underline{-2.6666666666666}492 \\
\hline
    \hline
\end{tabular}
}%
\caption{Benchmark point for the subprocess $d(p_{1})\bar{d}(p_{2})\rightarrow \nu_e(p_3)e^+(p_4)\bar{\nu}_\mu(p_5)\mu^-(p_6)b(p_7)\bar{b}(p_8)$.}
\label{tab:psp}
\end{center}
%}
\end{table}

\paragraph*{Partonic process: ${g}{g}\rightarrow{\nu_e}{e^+}{\bar{\nu}_\mu}{\mu^-}{b}{\bar{b}}$}
~\\
\begin{table}[ht]
%\TABLE{
%\centering
\begin{center}
{\footnotesize
\begin{tabular}{c c c c c}
\hline 
\hline
\textsc{particle} & $E$& $p_x$ & $p_y$ & $p_z$ \\ 
\hline
$p_1$ & 250.0000000000000000 & 0.0000000000000000 & 0.0000000000000000 & 250.0000000000000000 \\
$p_2$ & 250.0000000000000000 & 0.0000000000000000 & 0.0000000000000000 & -250.0000000000000000 \\
$p_3$ & 25.6649135887983881 & 15.7119189648809563 & -7.6256852614994211 & 18.8061776088085644 \\
$p_4$ & 154.5064489409292605 & -136.9768135536888565 & -46.2042967903236104 & -54.5413446032057010 \\
$p_5$ & 96.1603526761926730 & 22.3143342117606984 & 88.8440473781699325 & -29.2509685474019605 \\
$p_6$ & 61.0731578730670606 & 23.2528867537403734 & 2.3717946919774544 & 56.4234743714476039 \\
$p_7$ & 93.6199892430268648 & 58.9397665353324101 & -43.8226058551382351 & -57.9028973600646211 \\
$p_8$ & 68.9751376779857708 & 16.7579070879743455 & 6.4367458368138966 & 66.4655585304161036 \\
\hline
\hline
\end{tabular}

\vspace{0.3cm}

  \begin{tabular}{ r  r }
    \hline
    \hline
    \textsc{parameter} & \textsc{value} \\
    \hline
     ${m_W}$ \qquad & 80.376 GeV\\
    ${m_Z}$ \qquad & 91.1876 GeV \\
    ${m_t}$ \qquad & 171.2 GeV\\
    ${m_b}$ \qquad & 4.2 GeV\\
    ${N_f}$ \qquad & 4 \\
%    ${N_{f,\textrm{gen}}}$ \qquad & 2 \\
    $\mu$ \qquad & 500.0 GeV\\
    \hline
    \hline
\end{tabular}
\hspace{2.5cm}
  \begin{tabular}{ r  r }
    \hline
    \hline
    & ${g}{g}\rightarrow{\nu_e}{e^+}{\bar{\nu}_\mu}{\mu^-}{b}{\bar{b}}$ \\
    \hline
$a_0$     \qquad &   \underline{27.4387}494492212056 \\
$a_{-1}$  \qquad &   \underline{-7.955552394077}3458 \\
$a_{-2}$  \qquad &   \underline{-5.9999999999999}885 \\
\hline
    \hline
\end{tabular}
}%
\caption{Benchmark point for the subprocess $g(p_{1})g(p_{2})\rightarrow \nu_e(p_3)e^+(p_4)\bar{\nu}_\mu(p_5)\mu^-(p_6)b(p_7)\bar{b}(p_8)$.}
\label{tab:psp}
\end{center}
%}
\end{table}

\subsection{${p}\,{p}\rightarrow{W}\,{W}\,{b}\,{\bar{b}}+1$ jet}
\paragraph*{Partonic process: ${u}{\bar{u}}\rightarrow{\nu_e}{e^+}{\bar{\nu}_\mu}{\mu^-}{b}{\bar{b}}{g}$}
~\\
\begin{table}[ht]
%\TABLE{
%\centering
\begin{center}
{\footnotesize
\begin{tabular}{c c c c c}
\hline 
\hline
\textsc{particle} & $E$& $p_x$ & $p_y$ & $p_z$ \\ 
\hline
$p_1$ & 250.0000000000000000 & 0.0000000000000000 & 0.0000000000000000 & 250.0000000000000000 \\
$p_2$ & 250.0000000000000000 & 0.0000000000000000 & 0.0000000000000000 & -250.0000000000000000 \\
$p_3$ & 18.1435357791203664 & 10.6794732394995044 & -3.3652659935279523 & 14.2762644514536543 \\
$p_4$ & 132.6867460765342059 & -121.1435093670056773 & -5.2844663256145745 & -53.8711159925087628 \\
$p_5$ & 92.9820970636741606 & 17.2946496492191066 & 87.0024013115393018 & -27.8773677719967203 \\
$p_6$ & 46.6813566690488102 & 14.7875289839330328 & 9.6031823094958089 & 43.2233378690595345 \\
$p_7$ & 69.5794003816978091 & 41.1619960772432023 & -24.4397402957503331 & -50.4943772185567497 \\
$p_8$ & 53.6926389028252160 & 9.1521313351918430 & 14.4565277002773502 & 50.8934845655482206 \\
$p_9$ & 86.2342251270994211 & 28.0677300819189206 & -77.9726387064195876 & 23.8497740970007897 \\
\hline
\hline
\end{tabular}

\vspace{0.5cm}

  \begin{tabular}{ r  r }
    \hline
    \hline
    \textsc{parameter} & \textsc{value} \\
    \hline
%    ${m_H}$ \qquad & 125.0 \\
    ${m_W}$ \qquad & 80.376 GeV \\
    ${m_Z}$ \qquad & 91.1876 GeV  \\
    ${m_t}$ \qquad & 171.2 GeV \\
    ${m_b}$ \qquad & 0.0 GeV \\
    ${N_f}$ \qquad & 5 \\
  %  ${N_{f,\textrm{gen}}}$ \qquad & 5 \\
    $\mu^2$ \qquad & 500.0 GeV \\
    \hline
    \hline
\end{tabular}
\hspace{2.5cm}
  \begin{tabular}{ r  r }
    \hline
    \hline
    & ${u}{\bar{u}}\rightarrow{\nu_e}{e^+}{\bar{\nu}_\mu}{\mu^-}{b}{\bar{b}}{g}$ \\
    \hline
$a_0$     \qquad &   \underline{-38.5591246673}060795 \\
$a_{-1}$  \qquad &   \underline{-32.4828496060}584442 \\
$a_{-2}$  \qquad &   \underline{-8.333333333333}4405 \\
\hline
    \hline
\end{tabular}
}%
\caption{Benchmark point for the subprocess $u(p_{1})\bar{u}(p_{2})\rightarrow \nu_e(p_3)e^+(p_4)\bar{\nu}_\mu(p_5)\mu^-(p_6)b(p_7)\bar{b}(p_8)g(p_9)$.}
\label{tab:psp}
\end{center}
%}
\end{table}

\subsection{$p\,p\rightarrow H+3$ jets}
\paragraph*{Partonic process: ${g}{g}\rightarrow{H}{g}{g}{g}$}
~\\
\begin{table}[ht]
%\TABLE{
%\centering
\begin{center}
{\footnotesize
\begin{tabular}{c c c c c}
\hline 
\hline
\textsc{particle} & $E$& $p_x$ & $p_y$ & $p_z$ \\ 
\hline
$p_1$ & 250.0000000000000000 & 0.0000000000000000 & 0.0000000000000000 & 250.0000000000000000 \\
$p_2$ & 250.0000000000000000 & 0.0000000000000000 & 0.0000000000000000 & -250.0000000000000000 \\
$p_3$ & 144.2726812297522656 & 26.1871426153409317 & -66.9976759414063139 & 3.8924965402436307 \\
$p_4$ & 135.5632052379070274 & 5.8338195550562180 & 26.5129120011233681 & -132.8172227574218596 \\
$p_5$ & 75.6651361325424006 & -19.2292152047334604 & -65.8505932559228597 & 31.9241206051040152 \\
$p_6$ & 144.4989773997982923 & -12.7917469656637373 & 106.3353571962057913 & 97.0006056120742102 \\
\hline
\hline
\end{tabular}

\vspace{0.5cm}

  \begin{tabular}{ r  r }
    \hline
    \hline
    \textsc{parameter} & \textsc{value} \\
    \hline
    ${m_H}$ \qquad & 125.0 GeV\\
%    ${m_W}$ \qquad & 80.376 \\
%    ${m_Z}$ \qquad & 91.1876 \\
%    ${m_t}$ \qquad & 171.2 \\
%    ${m_b}$ \qquad & 4.2 \\
    ${N_f}$ \qquad & 5 \\
%    ${N_{f,\textrm{gen}}}$ \qquad & 1 \\
    $\mu$ \qquad & 500.0 GeV\\
    \hline
    \hline
\end{tabular}
\hspace{2.5cm}
  \begin{tabular}{ r  r }
    \hline
    \hline
    & ${g}{g}\rightarrow{H}{g}{g}{g}$ \\
    \hline
$a_0$     \qquad &   \underline{23.430792757}8718953 \\
$a_{-1}$  \qquad &   \underline{-56.37349644245}17315 \\
$a_{-2}$  \qquad &   \underline{-15.000000000000}8757 \\
\hline
    \hline
\end{tabular}
}%
\caption{Benchmark point for the subprocess $g(p_{1})g(p_{2})\rightarrow H(p_3)g(p_4)g(p_5)g(p_6)$.}
\label{tab:pgghgggpsp}
\end{center}
%}
\end{table}

\newpage

\subsection{$p\,p\rightarrow Z\,t\,\bar{t}+1$ jet}
\paragraph*{Partonic process: ${u}{\bar{u}}\rightarrow{t}{\bar{t}}{e^+}{e^-}{g}$}
~\\
\begin{table}[ht]
%\TABLE{
%\centering
\begin{center}
{\footnotesize
\begin{tabular}{c c c c c}
\hline 
\hline
\textsc{particle} & $E$& $p_x$ & $p_y$ & $p_z$ \\ 
\hline
$p_1$ & 250.0000000000000000 & 0.0000000000000000 & 0.0000000000000000 & 250.0000000000000000 \\
$p_2$ & 250.0000000000000000 & 0.0000000000000000 & 0.0000000000000000 & -250.0000000000000000 \\
$p_3$ & 183.2414081421947287 & -30.2337217736484156 & 32.1314578860740667 & 48.1815850690226029 \\
$p_4$ & 199.0327070603159996 & 74.5268539046026035 & -40.9270527537185629 & -55.4554134393922311 \\
$p_5$ & 70.1181125436057044 & -63.0760999348447697 & 21.5315800178266556 & 21.7794946135846281 \\
$p_6$ & 20.7607087314536756 & -7.2430664321972609 & -7.1983324871256098 & -18.0756472939650585 \\
$p_7$ & 26.8470635224299627 & 26.0260342360878454 & -5.5376526630565506 & 3.5699810507501222 \\
\hline
\hline
\end{tabular}

\vspace{0.5cm}

  \begin{tabular}{ r  r }
    \hline
    \hline
    \textsc{parameter} & \textsc{value} \\
    \hline
%      ${m_H}$ \qquad & 125.0 GeV \\
%    ${m_W}$ \qquad & 80.376 \\
    ${m_Z}$ \qquad & 91.1876 GeV\\
    ${m_t}$ \qquad & 171.2 GeV\\
  %  ${m_b}$ \qquad & 0.0 GeV\\
    ${N_f}$ \qquad & 5 \\
%    ${N_{f,\textrm{gen}}}$ \qquad & 2 \\
    $\mu$ \qquad & 500.0 GeV\\
    \hline
    \hline
\end{tabular}
\hspace{2.5cm}
  \begin{tabular}{ r  r }
    \hline
    \hline
    & ${u}{\bar{u}}\rightarrow{t}{\bar{t}}{e^+}{e^-}{g}$ \\
    \hline
$a_0$     \qquad &   \underline{-20.436776371}0913373 \\
$a_{-1}$  \qquad &   \underline{-25.907854281}5554513 \\
$a_{-2}$  \qquad &   \underline{-5.6666666665}792098 \\
\hline
    \hline
\end{tabular}
}%
\caption{Benchmark point for the subprocess $u(p_{1})\bar{u}(p_{2})\rightarrow t(p_3)\bar{t}(p_4)e^+(p_5)e^-(p_6)g(p_7)$.}
\label{tab:uutteegpsp}
\end{center}
%}
\end{table}

\paragraph*{Partonic process: ${g}{g}\rightarrow{t}{\bar{t}}{e^+}{e^-}{g}$}
~\\
\begin{table}[ht]
%\TABLE{
%\centering
\begin{center}
{\footnotesize
\begin{tabular}{c c c c c}
\hline 
\hline
\textsc{particle} & $E$& $p_x$ & $p_y$ & $p_z$ \\ 
\hline
$p_1$ & 250.0000000000000000 & 0.0000000000000000 & 0.0000000000000000 & 250.0000000000000000 \\
$p_2$ & 250.0000000000000000 & 0.0000000000000000 & 0.0000000000000000 & -250.0000000000000000 \\
$p_3$ & 174.2203895522303014 & -25.0977827305029138 & -19.5610151031829993 & 5.5472629175473589 \\
$p_4$ & 186.7123996976260685 & -14.0800163072181022 & 56.3619207264196902 & -46.6601246640355427 \\
$p_5$ & 60.3016377245591073 & 38.1795332240129639 & 22.1553968884492853 & 41.0822241824339116 \\
$p_6$ & 18.6184873501163182 & 5.2347824612577458 & -1.6661313271933778 & -17.7895792583830961 \\
$p_7$ & 60.1470856754682259 & -4.2365166475497116 & -57.2901711844925998 & 17.8202168224373914 \\
\hline
\hline
\end{tabular}

\vspace{0.5cm}

  \begin{tabular}{ r  r }
    \hline
    \hline
    \textsc{parameter} & \textsc{value} \\
    \hline
%    ${m_H}$ \qquad & 125.0 GeV \\
%    ${m_W}$ \qquad & 80.376 \\
    ${m_Z}$ \qquad & 91.1876 GeV\\
    ${m_t}$ \qquad & 171.2 GeV\\
 %   ${m_b}$ \qquad & 0.0 GeV\\
    ${N_f}$ \qquad & 5 \\
%    ${N_{f,\textrm{gen}}}$ \qquad & 2 \\
    $\mu$ \qquad & 500.0 GeV\\
    \hline
    \hline
\end{tabular}
\hspace{2.5cm}
  \begin{tabular}{ r  r }
    \hline
    \hline
    & ${g}{g}\rightarrow{t}{\bar{t}}{e^+}{e^-}{g}$ \\
    \hline
$a_0$     \qquad &   \underline{9.282642532}3344441 \\
$a_{-1}$  \qquad &   \underline{-26.2816094048}822784 \\
$a_{-2}$  \qquad &   \underline{-9.000000000000}5702 \\
\hline
    \hline
\end{tabular}
}%
\caption{Benchmark point for the subprocess $g(p_{1})g(p_{2})\rightarrow t(p_3)\bar{t}(p_4)e^+(p_5)e^-(p_6)g(p_7)$.}
\label{tab:ggtteegpsp}
\end{center}
%}
\end{table}

\newpage

\subsection{$p\,p\rightarrow H\,t\,\bar{t}+1$ jet}
\paragraph*{Partonic process: ${g}{g}\rightarrow{H}{t}{\bar{t}}{g}$}
~\\
\begin{table}[ht]
%\TABLE{
%\centering
\begin{center}
{\footnotesize
\begin{tabular}{c c c c c}
\hline 
\hline
\textsc{particle} & $E$& $p_x$ & $p_y$ & $p_z$ \\ 
\hline
$p_1$ & 250.0000000000000000 & 0.0000000000000000 & 0.0000000000000000 & 250.0000000000000000 \\
$p_2$ & 250.0000000000000000 & 0.0000000000000000 & 0.0000000000000000 & -250.0000000000000000 \\
$p_3$ & 126.3747891763443505 & 6.8355633672742222 & -3.2652801590882752 & 6.0992096455298030 \\
$p_4$ & 177.2234233286846745 & -31.9178657717747534 & -19.5439094615872051 & -15.8485716665707326 \\
$p_5$ & 174.8995128490773538 & 13.4406996200208031 & 24.1748981179500326 & -8.2771667589629576 \\
$p_6$ & 21.5022746458936318 & 11.6416027844796517 & -1.3657084972745175 & 18.0265287800038720 \\
\hline
\hline
\end{tabular}

\vspace{0.5cm}

  \begin{tabular}{ r  r }
    \hline
    \hline
    \textsc{parameter} & \textsc{value} \\
    \hline
    ${m_H}$ \qquad & 126.0 GeV\\
%    ${m_W}$ \qquad & 80.376 \\
%    ${m_Z}$ \qquad & 91.1876 \\
    ${m_t}$ \qquad & 172.5 GeV\\
  %  ${m_b}$ \qquad & 0.0 GeV\\
    ${N_f}$ \qquad & 5 \\
%    ${N_{f,\textrm{gen}}}$ \qquad & 1 \\
    $\mu$ \qquad & 500.0 GeV\\
    \hline
    \hline
\end{tabular}
\hspace{2.5cm}
  \begin{tabular}{ r  r }
    \hline
    \hline
    & ${g}{g}\rightarrow{H}{t}{\bar{t}}{g}$ \\
    \hline
$a_0$     \qquad &  \underline{-45.697933}4407767297 \\
$a_{-1}$  \qquad &  \underline{-35.92174974455}15619 \\
$a_{-2}$  \qquad &   \underline{-8.999999999999}0887 \\
\hline
    \hline
\end{tabular}
}%
\caption{Benchmark point for the subprocess $g(p_{1})g(p_{2})\rightarrow H(p_3)t(p_4)\bar{t}(p_5)g(p_6)$.}
\label{tab:ggttbarhgpsp}
\end{center}
%}
\end{table}

\subsection{$p\,p\rightarrow H+3$ jets (VBF)}
\paragraph*{Partonic process: ${u}{u}\rightarrow{g}{H}{u}{u}$ (VBF)}
~\\
\begin{table}[ht]
%\TABLE{
%\centering
\begin{center}
{\footnotesize
\begin{tabular}{c c c c c}
\hline 
\hline
\textsc{particle} & $E$& $p_x$ & $p_y$ & $p_z$ \\ 
\hline
$p_1$ & 250.0000000000000000 & 0.0000000000000000 & 0.0000000000000000 & 250.0000000000000000 \\
$p_2$ & 250.0000000000000000 & 0.0000000000000000 & 0.0000000000000000 & -250.0000000000000000 \\
$p_3$ & 40.1603071333660182 & -14.0012702315405289 & 0.4395613413056457 & 37.6380324509251736 \\
$p_4$ & 127.7084583092647421 & -23.4171446211731009 & -10.8486559189339324 & 4.2888607196847408 \\
$p_5$ & 145.0573545181100599 & -109.8833468186949176 & -94.5094823127907233 & 5.9366610719649477 \\
$p_6$ & 187.0738800392591656 & 147.3017616714085705 & 104.9185768904190468 & -47.8635542425748710 \\
\hline
\hline
\end{tabular}

\vspace{0.5cm}

  \begin{tabular}{ r  r }
    \hline
    \hline
    \textsc{parameter} & \textsc{value} \\
    \hline
    ${m_H}$ \qquad & 125.0 GeV\\
%    ${m_W}$ \qquad & 80.376 \\
    ${m_Z}$ \qquad & 91.1876 GeV\\
    ${w_Z}$ \qquad &  2.4952 GeV\\    
%    ${m_t}$ \qquad & 171.2 \\
%    ${m_b}$ \qquad & 0.0 \\
    ${N_f}$ \qquad & 5 \\
%    ${N_{f,\textrm{gen}}}$ \qquad & 1 \\
    $\mu$ \qquad & 500.0 GeV\\
    \hline
    \hline
\end{tabular}
\hspace{2.5cm}
  \begin{tabular}{ r  r }
    \hline
    \hline
    & ${u}{u}\rightarrow{g}{H}{u}{u}$ \\
    \hline
$a_0$     \qquad &   \underline{-94.68182872598}62359 \\
$a_{-1}$  \qquad &   \underline{-40.89041077795}83796 \\
$a_{-2}$  \qquad &   \underline{-8.333333333333}6821 \\
\hline
    \hline
\end{tabular}
}%
\caption{Benchmark point for the subprocess $u(p_{1})u(p_{2})\rightarrow g(p_3)H(p_4)u(p_5)u(p_6)$.}
\label{tab:psp}
\end{center}
%}
\end{table}

\newpage
\subsection{$p\,p\rightarrow H+4$ jets (VBF)}
\paragraph*{Partonic process: ${u}{u}\rightarrow{g}{g}{H}{u}{u}$ (VBF)}
~\\
\begin{table}[ht]
%\TABLE{
%\centering
\begin{center}
{\footnotesize
\begin{tabular}{c c c c c}
\hline 
\hline
\textsc{particle} & $E$& $p_x$ & $p_y$ & $p_z$ \\ 
\hline
$p_1$ & 250.0000000000000000 & 0.0000000000000000 & 0.0000000000000000 & 250.0000000000000000 \\
$p_2$ & 250.0000000000000000 & 0.0000000000000000 & 0.0000000000000000 & -250.0000000000000000 \\
$p_3$ & 77.8540004794647871 & -42.9509851388533761 & 6.0812524321357140 & -64.6488718781321126 \\
$p_4$ & 179.6868495846460405 & 66.2917119993839208 & -68.8295351006103004 & 152.1685510599341171 \\
$p_5$ & 140.8511015083574875 & -29.7530986170501173 & 2.6554844463192953 & -57.6344889901617350 \\
$p_6$ & 81.8035523206006161 & 2.3404081739038114 & 78.3533883805051659 & -23.3899591949725902 \\
$p_7$ & 19.8044961069310155 & 4.0719635826157594 & -18.2605901583498813 & -6.4952309966676500 \\
\hline
\hline
\end{tabular}

\vspace{0.5cm}

  \begin{tabular}{ r  r }
    \hline
    \hline
    \textsc{parameter} & \textsc{value} \\
    \hline
    ${m_H}$ \qquad & 125.0 GeV\\
%    ${m_W}$ \qquad & 80.376 \\
    ${m_Z}$ \qquad & 91.1876 GeV\\
    ${w_Z}$ \qquad &  2.4952 GeV\\
%    ${m_t}$ \qquad & 171.2 \\
%    ${m_b}$ \qquad & 0.0 \\
    ${N_f}$ \qquad & 5 \\
%    ${N_{f,\textrm{gen}}}$ \qquad & 1 \\
    $\mu$ \qquad & 500.0 GeV\\
    \hline
    \hline
\end{tabular}
\hspace{2.5cm}
  \begin{tabular}{ r  r }
    \hline
    \hline
    & ${u}{u}\rightarrow{g}{g}{H}{u}{u}$ \\
    \hline
$a_0$     \qquad &   \underline{-85.21172204982}22565 \\
$a_{-1}$  \qquad &   \underline{-54.226321485445}0339 \\
$a_{-2}$  \qquad &   \underline{-11.3333333333333}464 \\
\hline
    \hline
\end{tabular}
}%
\caption{Benchmark point for the subprocess $u(p_{1})u(p_{2})\rightarrow g(p_3)g(p_4)H(p_5)u(p_6)u(p_7)$.}
\label{tab:psp}
\end{center}
%}
\end{table}

\paragraph*{Partonic process: ${u}{u}\rightarrow{u}{\bar{u}}{H}{u}{u}$}
~\\
\begin{table}[ht]
%\TABLE{
%\centering
\begin{center}
{\footnotesize
\begin{tabular}{c c c c c}
\hline 
\hline
\textsc{particle} & $E$& $p_x$ & $p_y$ & $p_z$ \\ 
\hline
$p_1$ & 250.0000000000000000 & 0.0000000000000000 & 0.0000000000000000 & 250.0000000000000000 \\
$p_2$ & 250.0000000000000000 & 0.0000000000000000 & 0.0000000000000000 & -250.0000000000000000 \\
$p_3$ & 30.3977507329956786 & 18.6382269943320686 & -8.3093459176077840 & 22.5298582897146815 \\
$p_4$ & 146.5123801227517504 & -137.7596903034009301 & -42.8693308104768249 & -25.5022691658345515 \\
$p_5$ & 158.9318930425956466 & 27.2202771824319179 & 93.2652344091314376 & -13.9642109273819610 \\
$p_6$ & 73.9640816926485911 & 30.8380670569460271 & 1.2506117519626909 & 67.2170735086997695 \\
$p_7$ & 90.1938944090082799 & 61.0631190696906998 & -43.3371694330094499 & -50.2804517051979616 \\
\hline
\hline
\end{tabular}

\vspace{0.5cm}

  \begin{tabular}{ r  r }
    \hline
    \hline
    \textsc{parameter} & \textsc{value} \\
    \hline
    ${m_H}$ \qquad & 125.0 GeV\\
%    ${m_W}$ \qquad & 80.376 \\
    ${m_Z}$ \qquad & 91.1876 GeV\\
    ${w_Z}$ \qquad &  2.4952 GeV\\
%    ${m_t}$ \qquad & 171.2 \\
%    ${m_b}$ \qquad & 0.0 \\
    ${N_f}$ \qquad & 5 \\
%    ${N_{f,\textrm{gen}}}$ \qquad & 1 \\
    $\mu$ \qquad & 500.0 GeV\\  
    \hline
    \hline
\end{tabular}
\hspace{2.5cm}
  \begin{tabular}{ r  r }
    \hline
    \hline
    & ${u}{u}\rightarrow{u}{\bar{u}}{H}{u}{u}$ \\
    \hline
$a_0$     \qquad &   \underline{-36.99099313798}02686 \\
$a_{-1}$  \qquad &   \underline{-35.202979728253}2968 \\
$a_{-2}$  \qquad &   \underline{-8.0000000000000}533 \\
\hline
    \hline
\end{tabular}
}%
\caption{Benchmark point for the subprocess $u(p_{1})u(p_{2})\rightarrow u(p_3)\bar{u}(p_4)H(p_5)u(p_6)u(p_7)$.}
\label{tab:psp}
\end{center}
%}
\end{table}

\newpage

\subsection{$p\,p\rightarrow H+5$ jets (VBF)}
\paragraph*{Partonic process: ${u}{u}\rightarrow{g}{g}{g}{H}{u}{u}$ (VBF)}
~\\
\begin{table}[ht]
%\TABLE{
%\centering
\begin{center}
{\footnotesize
\begin{tabular}{c c c c c}
\hline 
\hline
\textsc{particle} & $E$& $p_x$ & $p_y$ & $p_z$ \\ 
\hline
$p_1$ & 250.0000000000000000 & 0.0000000000000000 & 0.0000000000000000 & 250.0000000000000000 \\
$p_2$ & 250.0000000000000000 & 0.0000000000000000 & 0.0000000000000000 & -250.0000000000000000 \\
$p_3$ & 24.3265597158113103 & 9.0293044031657743 & 17.9817135111430346 & -13.6715452237514459 \\
$p_4$ & 74.2975116145394878 & 11.5035704263332619 & 29.2267551012052458 & -67.3319009520241991 \\
$p_5$ & 97.6019108689663568 & -72.1835660496625877 & 4.7127971252997813 & -65.5244636825316746 \\
$p_6$ & 136.6365204371290929 & 23.6483158828115840 & -47.1886774719599131 & -16.0786999325225715 \\
$p_7$ & 120.9034677660200998 & 23.6253257672831865 & -14.6581830826285433 & 117.6632065216435592 \\
$p_8$ & 46.2340295975336915 & 4.3770495700687935 & 9.9255948169403965 & 44.9434032691863266 \\
\hline
\hline
\end{tabular}

\vspace{0.5cm}

  \begin{tabular}{ r  r }
    \hline
    \hline
    \textsc{parameter} & \textsc{value} \\
    \hline
    ${m_H}$ \qquad & 125.0 GeV\\
%    ${m_W}$ \qquad & 80.376 \\
    ${m_Z}$ \qquad & 91.1876 GeV\\
    ${w_Z}$ \qquad &  2.4952 GeV\\
%    ${m_t}$ \qquad & 171.2 \\
%    ${m_b}$ \qquad & 0.0 \\
    ${N_f}$ \qquad & 5 \\
%    ${N_{f,\textrm{gen}}}$ \qquad & 1 \\
    $\mu$ \qquad & 500.0 GeV\\
    \hline
    \hline
\end{tabular}
\hspace{2.5cm}
  \begin{tabular}{ r  r }
    \hline
    \hline
    & ${u}{u}\rightarrow{g}{g}{g}{H}{u}{u}$ \\
    \hline
$a_0$     \qquad &   \underline{-164.8823178520}154897 \\
$a_{-1}$  \qquad &   \underline{-81.44727157941}69630 \\
$a_{-2}$  \qquad &   \underline{-14.3333333333333}570 \\
\hline
    \hline
\end{tabular}
}%
\caption{Benchmark point for the subprocess $u(p_{1})u(p_{2})\rightarrow g(p_3)g(p_4)g(p_5)H(p_6)u(p_7)u(p_8)$.}
\label{tab:psp}
\end{center}
%}
\end{table}

\section{Conclusions}

The integrand reduction techniques have changed the way 
to perform the decomposition of scattering amplitudes in terms of
independent integrals. In these approaches, the coefficients which multiply each integral can be
completely determined algebraically by relying on the knowledge of the universal structure of the residues of amplitudes at each multiple cuts. The residues are irreducible polynomials in the
components of the loop momenta which are not constrained by the on-shell
conditions defining the cuts. The coefficients of the master integrals
are a subset of the coefficients of the residues.

The generalized unitarity strategy implemented within the integrand
decomposition requires to solve a triangular system, 
where the coefficients of the residues, hence of the master integrals, can be projected
out of cuts only after removing the contributions of higher-point
residues. By adding one more ingredient to this strategy, namely the Laurent
series expansion of the integrand with respect to the unconstrained
components of the loop momentum, 
we improved the system-solving strategy, that became diagonal.

We demonstrated that this novel reduction algorithm, 
now implemented in the computer code 
{\sc Ninja}, and currently interfaced to the {\sc GoSam} framework, 
yields a very efficient and accurate evaluation of multi-particle
one-loop scattering amplitudes, no matter whether massive particles go
around the loop or participate to the scattering as external legs.
We used  {\Gosam}+{\Ninja} to compute NLO corrections to a set of non-trivial
processes involving up to {\it eight} particles.

The level of automation reached in less than a decade by the evaluation of scattering amplitudes at
next-to-leading order has been heavily stimulated by
the awareness that the methods for computing the virtual
contributions were simply not sufficient, while the techniques for 
controlling the infrared divergencies and, finally, for performing  
phase-space integration were already available.
Nowadays, the scenario is changed, and one-loop contributions to many 
multi-particle scattering reactions are
waiting to be integrated. We hope that these advancements can stimulate the
developments of novel methods for computing cross sections and
distributions at next-to-leading-order accuracy for {\it high-multiplicity} final states.

\section*{Acknowledgments}
We thank all the other members of the {\Gosam} project for collaboration on
the common development of the code. We also would like to thank
Valery Yundin for comparisons of $Z+3$ jets and $W+3$ jets with {\njet}.
% for providing us data to compare our results for $Z$+ jets and $W$+jets with {\njet}.
The work of H.v.D., G.L., P.M., and T.P. was supported by the Alexander von Humboldt
Foundation, in the framework of the Sofja Kovalevskaja Award Project ÒAdvanced Mathematical Methods for
Particle PhysicsÓ, endowed by the German Federal Ministry of Education and Research. The work 
of G.O. was supported in part by the National Science Foundation under Grant PHY-1068550.
G.O. also wishes to acknowledge the kind hospitality of the Max Planck Institut f\"ur Physik 
in Munich at several stages during the completion of  this project.

\bibliographystyle{utphys}
\bibliography{references}

%%%%%%%%%%%%%%%%%%%%%%%%%%%%%%%%%%%%%%%%%%%%%%%%%%%%%%%%%%%%%%%%%%%%%%%
%\begin{thebibliography}{999}
%
%\end{thebibliography}
%%%%%%%%%%%%%%%%%%%%%%%%%%%%%%%%%%%%%%%%%%%%%%%%%%%%%%%%%%%%%%%%%%%%%%%

\end{document}